\documentstyle[preprint,aps,prb,epsfig]{revtex}
\begin{document}
\draft
\pagestyle{plain}
\newcommand{\D}{\displaystyle}
\title{\bf Effect of anisotropic impurity scattering in superconductors} 
\author{Grzegorz Hara\'n\cite{AA} and A. D. S. Nagi}
\vspace{0.4cm}
\address{Department of Physics, University of Waterloo, 
Waterloo, Ontario, Canada, N2L 3G1}
\date{\today} 
\maketitle

\begin{abstract}
We discuss the weak-coupling BCS theory of a superconductor with the impurities, 
accounting for their anisotropic momentum-dependent potential. The impurity 
scattering process is considered in the t-matrix approximation and its influence  
on the superconducting critical temperature $T_c$ is studied in the Born and unitary 
limit for a $d_{x^2-y^2}$- and $(d_{x^2-y^2}+s)$-wave superconductors. 
We observe a significant dependence of the pair-breaking strength on  
the symmetry of the scattering potential and classify the impurity potentials 
according to their ability to alter $T_c$. A good agreement with the 
experimental data for Zn doping and oxygen irradiation in the overdoped cuprates 
is found.  
\end{abstract}

\vspace{0.5cm}

\section{\bf Introduction}
Several experiments probing the effect of impurities or lattice defects on  
superconductivity in the cuprates have been carried out  
\cite{I1,I2,I3,I4,I5,I6,I7,I8,I9,I10,I11,I12,I13,I14,I15,I16,I17,I18,I19,I20,I21,I21a} 
in order to get more insight into the symmetry of the superconducting state. 
The most thoroughly studied defects are Zn and Ni substitutions 
\cite{I1,I2,I3,I4,I5,I6,I7,I8,I9,I10,I11,I12,I13,I14,I15,I16,I17,I18} 
on the planar Cu sites and irradiation induced oxygen vacancies \cite{I19,I20,I21,I21a} 
in the copper oxygen planes. 
Yet the results are difficult to explain within a standard Abrikosov-Gorkov type theory  
of impurity scattering \cite{II1} for the scenario of the $d$-wave superconductivity, 
which is predicted to be extremely suppressed by the impurities. \cite{I22,I22a,I22b,I22c} 
This issue was critically examined by Radtke et al. \cite{I22} who considered 
isotropic nonmagnetic impurity scattering in the Born approximation and obtained 
the critical temperature in both weak- and strong-coupling 
approach close to the Abrikosov-Gorkov scaling function. A comparison to the electron 
irradiation data \cite{I19} in $YBa_2Cu_3O_{7-\delta}$ ($Y\!-\!123$) showed that the 
theoretically predicted 
$T_c$ was about twice as much reduced by the impurities than observed. 
This inconsistency can be settled down within the weak impurity scattering 
model if the impurity 
scattering rate a factor of 3 less than the one deduced from the transport 
measurements is assumed \cite{I20,I21} which is equivalent to an introduction of 
two separate relaxation time scales - one defining the scattering time in pair-breaking 
processes and the other representing the transport scattering time. Such a distinction 
occurs naturally as a consequence of an impurity momentum-dependent scattering probability.   
\cite{II7} The issue of a possible anisotropy in the impurity scattering potential  
was suggested in a discussion of the irradiation 
data by Giapintzakis et al. \cite{I19} where the authors evoked a model by Millis et 
al. \cite{II4} A more general formulation of the problem was given in our previous  
paper, \cite{I24} where the effective correlation between two impurity vertex functions 
was assumed in the form $|w_0|^2+|w_1|^2f\left({\bf k}\right)f\left({\bf k'}\right)$,  
with $|w_0|$ and $|w_1|$ 
representing isotropic and anisotropic impurity scattering amplitude respectively 
and $f\left({\bf k}\right)$ determining the symmetry of the impurity potential. 
\cite{I24a} Our analysis showed that the symmetry of the anisotropic potential  
is an important factor and a significant reduction in the pair-breaking 
strength appears for large values of $\left<ef\right>^2=
\left[\int_{FS}dS_{k}n\left({\bf k}\right)e\left({\bf k}\right)
f\left({\bf k}\right)\right]^2$, where $\int_{FS}dS_{k}$ denotes integration over 
the Fermi surface (FS) and $n\left({\bf k}\right)$ is the normalized 
($\int_{FS}dS_{k}n\left({\bf k}\right)=1$) angle-resolved FS density of states. 
Particularly low $T_c$ suppression is predicted for  
$\left<ef\right>^2=1$ that is for the anisotropy of the scattering potential 
in phase with the order parameter, which in the case of the $d_{x^2-y^2}$-wave 
superconductor corresponds to $f\left({\bf k}\right)\sim k_x^2-k_y^2$. Within this approach 
we were also able to understand quantitatively the irradiation data \cite{I19} in $Y-123$.     

Through the postulation of the analytic form of the square value of the impurity potential 
this approach has been designed for the second-order Born approximation and cannot be efficiently 
extended to include multiple impurity scattering processes. Such a generalization is 
important from the theoretical and experimental points of view, especially in the light of 
the recent experiments suggesting a possible strong (close to unitary) scattering by Zn atoms 
in $Y\!-\!123$ and $La_{2-x}Sr_xCuO_4$ ($La\!-\!214$) compounds. \cite{I2}   
For that purpose a model based on the assumption of the impurity potential 
and not its square value is needed. 

In this paper we study in the t-matrix approximation the pair-breaking effect of 
the nonmagnetic impurities with the anisotropic momentum-dependent factorizable 
potential.  
The influence on the superconducting transition temperature is analyzed quantitatively 
in the Born and unitary scattering limits. In particular, we find that the 
scattering from impurity potential 
in phase with the order parameter leads to a stronger suppression of 
the critical temperature than from the impurity potential orthogonal to the superconducting 
state. On the other hand, the superconducting state 
appears very robust to the potential scattering with its maxima in the region 
of the nodes of the order parameter.  
Finally, we find that the Zn and electron irradiation $T_c$ suppression data in 
$Y\!-\!123$ and $La\!-\!214$ are in the range predicted by our model.  

This work is organized as follows. In Sec. {\bf II}, we introduce the anisotropic  
momentum-dependent impurity potential. In Sec. {\bf III}, we derive the  
expressions necessary for the analysis of the scattering process of arbitrary 
strength. In Sec. {\bf IV}, we obtain the self-energies due to impurity scattering 
at the superconducting-normal state phase transition which allows the evaluation 
of the critical temperature. Then we examine analytically Born and unitary scattering 
limits. In Sec. {\bf V}, we calculate numerically the critical temperature for 
$d$- and $(d+s)$-wave superconducting states in the presence of impurities  
accounting for their anisotropic potential of $p$-, $d$-, $f$-, and $g$-wave symmetry 
in Born and unitary limit. In Sec. {\bf VI}, we compare the results with data on 
Zn substituted and irradiated overdoped $La\!-\!214$ and optimally doped $Y\!-\!123$ 
samples. In Sec. {\bf VII}, we present our conclusions.  
Except for comparing to the experimental data we assume $\hbar=k_{B}=1$ in the 
calculations.

\section{\bf Impurity scattering potential} 
As the impurity scattering strength is rather impossible to be determined 
from first principles, the information about the scattering process is usually  
deduced from a comparison of the experimental data with the theoretical models. 
These models assume a certain phenomenological impurity potential 
\cite{II1,I22,II4,I24,I25,II2,II3,I26,I28} 
which is verified by a fit to the available data. We proceed 
in the same way by analyzing the impurity scattering potential of the factorizable 
form 

\begin{equation}
\label{eI1}
v\left({\bf k},{\bf k'}\right)=v_i+
v_af\left({\bf k}\right)f\left({\bf k'}\right)
\end{equation}

\noindent
The above interaction consists of two channels - the isotropic scattering channel  
with the scattering amplitude $v_i$, and the anisotropic one determined by the 
scattering strength $v_a$ and 
momentum-dependent function $f\left({\bf k}\right)$.  
We assume that the $f\left({\bf k}\right)$ average value over the Fermi surface  
$\left<f\right>\!=\!\int_{FS}dS_k n\left({\bf k}\right)f\left({\bf k}\right)
\!=\!0$. Therefore, 
$f\left({\bf k}\right)$ is orthogonal to the isotropic s-wave term in a sense of a   
scalar product defined as the FS integral and in consequence a symmetry other   
than the identity is introduced into the impurity potential. An additional normalization 
$\left<f^2\right>\!=\!1$ gives $v_a$ the meaning of the scattering strength magnitude in  
the anisotropic channel. The chosen potential 
depends on the absolute orientations in the crystal of the incoming and 
outgoing (scattered) particles momenta, not only on the angle between ${\bf k}$ 
and ${\bf k'}$. The usually assumed isotropic conditions \cite{II5}  
of a spherical (cylindrical in two dimensions) constant energy surface and the 
scattering probability 
dependent only on the angle of deflexion are broken here. \cite{II5a} This leads to a 
momentum-dependence of the time between scattering events determined by the imaginary  
part of the self-energy and a momentum-dependent relaxation time in the Boltzmann   
equation. \cite{II6,II7} This second quantity is worth mentioning as it provides 
the information about the normal state transport properties and can be used as an additional physical 
assessment of the phenomenological model. \cite{II7} A constraint on the potential (1) which 
follows immediately from the analysis of the normal state properties is a nonzero value of 
the scattering amplitude in the isotropic channel $v_i$.       
A lack of the s-wave scattering may lead in some cases to an infinite value of some elements 
in the dc conductivity tensor, \cite{II7} that is to nonphysical transport properties.  

The anisotropic potential from Eq. (\ref{eI1}) distinguishes a certain set of 
coordinates which we think should coincide with the main directions in the crystal. 
This assumption seems rather plausible because 
the impurity potential is determined by the dopant atom itself as well as its 
substitution site in the crystal. The potential produced by a given 
sort of impurity (defect)   
may be considered unique since the impurities tend to selectively substitute 
at characteristic sites in the crystal. The Zn and Ni atoms occupy the 
in-plane Cu sites 
\cite{I1,I2,I3,I4,I5,I6,I7,I8,I9,I10,I11,I12,I13,I14,I15,I16,I17,I18}  
and the electron irradiation displaces the oxygen atoms from the $CuO_2$ 
planes. \cite{I19,I20,I21,I21a} Therefore on a short length scale of the order of 
magnitude of the lattice constant a given impurity dopant produces the same potential 
of the same orientation throughout the crystal.  
On the other hand, on a large length scale the impurity distribution 
in the system is random and the Abrikosov-Gorkov's method \cite{II1}
of averaging the Green's functions is applicable.  

We have not made any assumption about the electron energy band yet and we use a general 
formalism deriving the equations valid for an arbitrary Fermi surface.  
For the computational simplicity, however, the numerical results are 
obtained for a cylindrical FS. This approximation allows to study   
the effect of the anisotropy of the impurity potential alone. The anisotropy of 
the Fermi surface may equally enhance or suppress the pair-breaking effect of the 
impurities. The detailed quantitative calculations are needed to answer this 
question which is beyond the scope of the present paper.  

For the numerical calculations we take the function $f\left({\bf k}\right)$ in  
the scattering potential (\ref{eI1}) proportional to  
the harmonic functions which in a polar angle notation read $sin(l\phi)$ and $cos(l\phi)$, 
where $l$ is an integer number. Therefore we study the effect of the basis elements 
in a space of the functions determined by a two-dimensional momentum vector. 

\section{\bf T-matrix approximation for the self-energy}
We study the effect of potential scattering by spinless, noninteracting 
impurities on the single-particle propagator for superconducting electrons 

\begin{equation}
\label{eII1}
\hat{G}\left({\bf k},\omega\right)=\left[i\omega\hat{\tau}_0\hat{\sigma}_0
-\xi_k\hat{\tau}_3\hat{\sigma}_0-\Delta\left({\bf k}\right)i\hat{\tau}_2
\hat{\sigma}_2-\hat{\Sigma}\left({\bf k},\omega\right)\right]^{-1}
\end{equation}

\noindent
Here $\xi_{k}$ is the quasiparticle energy, $\omega=\pi T(2m+1)$, where  
T is the temperature and $m$ is an integer. $\hat{\tau}_j$, $\hat{\sigma}_j$ 
($j=1,2,3$) are the Pauli matrices and $\hat{\tau}_0$, $\hat{\sigma}_0$ are the  
unit matrices in particle-hole (Nambu) and spin space respectively.  
The order parameter $\Delta\left({\bf k}\right)$ is defined as 

\begin{equation}
\label{eII2}
\Delta\left({\bf k}\right)=\Delta e\left({\bf k}\right)
\end{equation}

\noindent
where $e\left({\bf k}\right)$ is a momentum-dependent real function  
which may belong to a one-dimensional (1D) irreducible representation 
of the crystal point group or may be given by a linear combination 
of the basis functions of different 1D representations. We normalize  
$e\left({\bf k}\right)$ by taking its average value over the Fermi surface 
$\left<e^{2}\right>\!=\!1$. 
This normalization gives $\Delta$ the meaning of the magnitude  
of the order parameter. The self-energy $\hat{\Sigma}\left({\bf k},\omega\right)$ 
and consequently the Green's function (Eq. (\ref{eII1})) have been obtained 
by applying Abrikosov-Gorkov's technique \cite{II1} of averaging over the 
coordinates of the impurities and depend on only one momentum vector ${\bf k}$. 
In this approximation  

\begin{equation}
\label{eII3}
\hat{\Sigma}\left({\bf k},\omega\right)=n\hat{T}
\left({\bf k},{\bf k},\omega\right)
\end{equation}

\noindent
where $n$ is the impurity concentration and $\hat{T}$ obeys the 
Lippmann-Schwinger equation \cite{II2,II3,III1,III2,III3,III4} 
 
\begin{equation}
\label{eII4}
\hat{T}\left({\bf k},{\bf k'},\omega\right)=\hat{v}\left({\bf k},{\bf k'}\right)
+\sum_{{\bf k''}}\hat{v}\left({\bf k},{\bf k''}\right)
\hat{G}\left({\bf k''},\omega\right)\hat{T}\left({\bf k''},{\bf k'},\omega\right)
\end{equation}

\noindent
Since the scattering potential for a single electron 

\begin{equation}
\label{eII5}
\hat{v}\left({\bf k},{\bf k'}\right)=
v\left({\bf k},{\bf k'}\right)\hat{\tau}_3\hat{\sigma}_0
\end{equation}

\noindent
is momentum-dependent (Eq. (\ref{eI1})), the vertex part 
$\hat{T}\left({\bf k},{\bf k'},\omega\right)$ is a function of two momenta 
${\bf k}$ and ${\bf k'}$. Therefore it should be evaluated from Eq. (\ref{eII4}) 
using the explicit form of $v\left({\bf k},{\bf k'}\right)$ first, and then 
the self-energy can be obtained according to Eq. (\ref{eII3}) 
by taking ${\bf k'}={\bf k}$. We proceed to a solution by defining   

\begin{equation}
\label{eII6}
\hat{g}_i\left(\omega\right)=\sum_{{\bf k}}f^{i}\left({\bf k}\right)
\hat{G}\left({\bf k},\omega\right)\;,\;\;\;i=0,1,2
\end{equation}

\noindent
and expanding all matrix quantities as  

\begin{equation}
\label{eII7}
\hat{\Sigma}\left({\bf k},\omega\right)=\Sigma_0\left({\bf k},\omega\right)
\hat{\tau}_0\hat{\sigma}_0+\Sigma_1\left({\bf k},\omega\right)
\hat{\tau}_1\hat{\sigma}_2+\Sigma_2\left({\bf k},\omega\right)
\hat{\tau}_2\hat{\sigma}_2+\Sigma_3\left({\bf k},\omega\right)
\hat{\tau}_3\hat{\sigma}_0
\end{equation}

\begin{equation}
\label{eII8}
\hat{g}_i\left(\omega\right)=g_{i0}\left(\omega\right)
\hat{\tau}_0\hat{\sigma}_0+g_{i1}\left(\omega\right)
\hat{\tau}_1\hat{\sigma}_2+g_{i2}\left(\omega\right)
\hat{\tau}_2\hat{\sigma}_2+g_{i3}\left(\omega\right)
\hat{\tau}_3\hat{\sigma}_0
\end{equation}

\noindent
The expression for the one-particle Green's function is then 

\begin{equation}
\label{eII8a}
\hat{G}\left({\bf k},\omega\right)=
-\frac{i\tilde{\omega}\hat{\tau}_0\hat{\sigma}_0 +\tilde{\xi_k}\hat{\tau}_3
\hat{\sigma}_0+\tilde{\Delta}\left({\bf k}\right)i\hat{\tau}_2
\hat{\sigma}_2+\tilde{\Delta}'\left({\bf k}\right)\hat{\tau}_1\hat{\sigma}_2}
{\tilde{\omega}^2+\tilde{\xi_k}^2-\tilde{\Delta}^2\left({\bf k}\right)
+\tilde{\Delta}'^2\left({\bf k}\right)}
\end{equation}

\noindent
with $\tilde{\omega}=\omega+i\Sigma_0\left({\bf k},\omega\right)$, 
$\tilde{\xi_k}=\xi_k+\Sigma_3\left({\bf k},\omega\right)$, 
$\tilde{\Delta}\left({\bf k}\right)=\Delta\left({\bf k}\right)-i 
\Sigma_2\left({\bf k},\omega\right)$ and $\tilde{\Delta}'\left({\bf k}\right)=
\Sigma_1\left({\bf k},\omega\right)$. 
The formal solutions for the self-energies $\Sigma_j\left({\bf k},\omega\right)$ 
$(j=0,1,2,3)$ obtained through Eqs. (\ref{eII3}) and (\ref{eII4}) read 

\begin{equation}
\label{eII9}
\begin{array}{l}
\Sigma_j\left({\bf k},\omega\right)=
s_j\left(v_i,v_a,\hat{g}_0,\hat{g}_1,\hat{g}_2\right)
u\left(v_i,v_a,\hat{g}_0,\hat{g}_1,\hat{g}_2\right)\\
+s_j\left(v_a,v_i,\hat{g}_2,\hat{g}_1,\hat{g}_0\right)
u\left(v_a,v_i,\hat{g}_2,\hat{g}_1,\hat{g}_0\right)
f^{2}\left({\bf k}\right)\\
+\left[t_{3-j}\left(v_i,v_a,\hat{g}_0,\hat{g}_1,\hat{g}_2\right)
+t_{3-j}\left(v_a,v_i,\hat{g}_2,\hat{g}_1,\hat{g}_0\right)
\right]f\left({\bf k}\right)
\end{array}
\end{equation}

\noindent
Note a permutation: $v_i\leftrightarrow v_a$, $\hat{g}_0\leftrightarrow\hat{g}_2$,  
in the arguments of the anisotropic parts (proportional to $f\left({\bf k}\right)$ 
and $f^2\left({\bf k}\right)$) of $\Sigma_j\left({\bf k},\omega\right)$ in 
Eq. (\ref{eII9}). 
The functions $s_j$, $t_j$ and $u$ are given by a series of equations   

\begin{equation}
\label{eII10}
\begin{array}{l}
s_0\left(v_i,v_a,\hat{g}_0,\hat{g}_1,\hat{g}_2\right)=
a_0c_3+a_1c_2+a_2c_1+a_3c_0\\
s_1\left(v_i,v_a,\hat{g}_0,\hat{g}_1,\hat{g}_2\right)=
a_0c_2+a_1c_3+ia_2c_0-ia_3c_1\\
s_2\left(v_i,v_a,\hat{g}_0,\hat{g}_1,\hat{g}_2\right)=
a_0c_1-ia_1c_0+a_2c_3+ia_3c_2\\
s_3\left(v_i,v_a,\hat{g}_0,\hat{g}_1,\hat{g}_2\right)=
a_0c_0+ia_1c_1-ia_2c_2+a_3c_3
\end{array}
\end{equation}

\begin{equation}
\label{eII11}
t_j\left(v_i,v_a,\hat{g}_0,\hat{g}_1,\hat{g}_2\right)=
v_in c_j\left(c^2_0+c^2_1+c^2_2-c^2_3\right)^{-1}
\end{equation}

\begin{equation}
\label{eII12}
u\left(v_i,v_a,\hat{g}_0,\hat{g}_1,\hat{g}_2\right)=v_iv^{-1}_an
d^{-1}\left(c^2_0+c^2_1+c^2_2-c^2_3\right)^{-1}
\end{equation}

\noindent
with $d=-g^2_{10}+g^2_{11}+g^2_{12}+g^2_{13}$ and 
the coefficients $a_j$, $c_j$ $(j=0,1,2,3)$ determined by twelve 
$g_{ik}$ elements (Eqs. (\ref{eII6}) and (\ref{eII8})) given by the integrals of  
the products of the Green's function $\hat{G}\left({\bf k},\omega\right)$ and  
appropriate powers of the impurity potential anisotropy function $f\left({\bf k}\right)$.  
These coefficients are introduced in order to shorten and simplify the 
notation of the self-energy functions. We define them in the sequential formulas 
with a use of additional $b_j$ parameters 

\begin{equation}
\label{eII16}
\begin{array}{l}
a_0=g_{13}\left(1-v_ag_{23}\right)-v_a\left(g_{12}g_{22}+g_{11}g_{21}
-g_{10}g_{20}\right)\\
a_1=ig_{12}\left(1-v_ag_{23}\right)+v_a\left(ig_{13}g_{22}-g_{11}g_{20}
+g_{10}g_{21}\right)\\
a_2=-ig_{11}\left(1-v_ag_{23}\right)-v_a\left(ig_{13}g_{21}+g_{12}g_{20}
-g_{10}g_{22}\right)\\
a_3=-g_{10}\left(1-v_ag_{23}\right)-v_a\left(g_{13}g_{20}-ig_{12}g_{21}
+ig_{11}g_{22}\right)
\end{array}
\end{equation}

\begin{equation}
\label{eII15}
\begin{array}{l}
b_0=\left(1-v_ig_{03}\right)a_0+v_i\left(ig_{02}a_1-ig_{01}a_2
-g_{00}a_3\right)\\
b_1=\left(1-v_ig_{03}\right)a_1+v_i\left(ig_{02}a_0+ig_{00}a_2
+g_{01}a_3\right)\\
b_2=\left(1-v_ig_{03}\right)a_2+v_i\left(g_{02}a_3-ig_{01}a_0
-ig_{00}a_1\right)\\
b_3=\left(1-v_ig_{03}\right)a_3-v_i\left(g_{02}a_2+g_{01}a_1
+g_{00}a_0\right)
\end{array}
\end{equation}

\begin{equation}
\label{eII14}
\begin{array}{l}
c_0=v_a^{-1}d^{-1}b_0-v_ig_{13}\\
c_1=iv_a^{-1}d^{-1}b_1-v_ig_{12}\\
c_2=-iv_a^{-1}d^{-1}b_2-v_ig_{11}\\
c_3=-v_a^{-1}d^{-1}b_3+v_ig_{10}
\end{array}
\end{equation}

\noindent
The self-energies can be evaluated with a simultaneous solution of the gap 
equation 

\begin{equation}
\label{eII17}
\Delta\left({\bf k}\right)i\hat{\sigma}_2=-T\sum_{\omega}\sum_{{\bf k'}}
\frac{1}{2}V_{{\bf k},{\bf k'}}tr\left[\left(\hat{\tau}_1+\hat{\tau}_2\right)\hat{G}
\left({\bf k'},\omega\right)\right]
\end{equation}

\noindent
where $V_{{\bf k},{\bf k'}}=-V_0e\left({\bf k}\right)e\left({\bf k'}\right)$,  
$V_0>0$, is the pair potential. 
The above lengthy expressions have been derived without any additional constraints  
and are fundamental to the considerations within the model. 
This generally complicated problem simplifies    
with the assumption of particle-hole symmetry of the excitation spectrum. 
It has been shown by Hirschfeld et al. \cite{III2} that in this case 
the $\hat{\tau}_3$ component of the integrated Green's function ($g_{03}$) may  
be neglected in the presence of $s$-wave scatterers. Inclusion of higher-order 
angular momentum waves in the scattering potential makes this analysis considerably 
more difficult. In this paper, however, we calculate the self-energies at the 
phase transition and for that purpose we need to show only the consistency of 
the assumption $\Sigma_3=0$ at the critical temperature $T_c$. 
In the following we assume a particle-hole 
symmetry of the energy spectrum, take $g_{i3}=0$ $(i=0,1,2)$ and check if this  
condition leads to a vanishing self-energy $\Sigma_3$. 

\section{Self-energy at phase transition}
We consider the effect of anisotropic impurity scattering 
on the critical temperature. At the superconducting-normal state phase 
transition the gap equation (\ref{eII17}) transforms into  

\begin{equation}
\label{eIII1}
1=V_0T_c\sum_{\omega}\sum_{{\bf k}}e\left({\bf k}\right)
\frac{\D e\left({\bf k}\right)+\frac{1}{i}
\left(\Sigma_1\left({\bf k},\omega\right)/\Delta\right)_{\Delta=0}
+\frac{1}{i}
\left(\Sigma_2\left({\bf k},\omega\right)/\Delta\right)_{\Delta=0}}
{\D\left(\omega+i\Sigma_0\left({\bf k},\omega\right)_{\Delta=0}\right)^2
+\left(\xi_k+\Sigma_3\left({\bf k},\omega\right)_{\Delta=0}\right)^2}
\end{equation}

\noindent
In order to find $T_c$ 
the self-energies in $\Delta\rightarrow 0$ limit are to be obtained. 
For the sake of convenience we introduce new parameters $c_i=1/(\pi N_0 v_i)$ 
and $c_a=1/(\pi N_0 v_a)$ describing the scattering strength in the 
isotropic ($c_i$) and anisotropic ($c_a$) channel respectively. 
$N_0$ represents the overall single-spin density of states at the Fermi level. 
A new 
measure of impurity concentration $\Gamma=n/(\pi N_0)$ is also used.   
Taking into account that $\left(g_{00}\right)_{\Delta=0}=-i\pi N_0 sgn(\omega)$ 
and $\left(g_{20}\right)_{\Delta=0}=\left(g_{00}\right)_{\Delta=0}$ 
(because of normalization $\left<f^2\right>=1$) we obtain   

\begin{equation}
\label{eIII2}
\Sigma_0\left({\bf k},\omega\right)_{\Delta=0}=
-i\Gamma\left[\frac{1}{c^2_i+1}+\frac{1}{c^2_a+1}
f^{2}\left({\bf k}\right)\right]sgn\left(\omega\right) 
\end{equation}

\begin{equation}
\label{eIII4}
\Sigma_3\left({\bf k},\omega\right)_{\Delta=0}=
\Gamma\left[\frac{c_i}{c^2_i+1}+\frac{c_a}{c^2_a+1}
f^{2}\left({\bf k}\right)\right]
\end{equation}

\noindent
Calculation of    
$\left(\Sigma_1\left({\bf k},\omega\right)/\Delta\right)_{\Delta=0}$ and 
$\left(\Sigma_2\left({\bf k},\omega\right)/\Delta\right)_{\Delta=0}$ 
quantities is more tedious and requires a solution of two sets of 
two linear equations which determine four unknowns $\left(g_{j1}/\Delta\right)
_{\Delta=0}$, $\left(g_{j2}/\Delta\right)_{\Delta=0}$ $(j\!=\!0,2)$ (note 
that $g_{11}\!=\!g_{12}\!=\!0$ because of $\left<f\right>\!=\!0$). This procedure 
leads to    
$\left(\Sigma_1\left({\bf k},\omega\right)/\Delta\right)_{\Delta=0}=0$ and 

\begin{equation}
\label{eIII5}
\begin{array}{l}
\D\left(\Sigma_2\left({\bf k},\omega\right)/\Delta\right)_{\Delta=0}= 
-\frac{\Gamma}{\pi N_0}\left[\frac{1}{c^2_i+1}\left(g_{02}/\Delta\right)
_{\Delta=0}
+\frac{1}{c^2_a+1}\left(g_{22}/\Delta\right)
_{\Delta=0}f^{2}\left({\bf k}\right)\right]sgn\left(\omega\right)
\end{array}
\end{equation}

\noindent
where 

\begin{equation}
\label{eIII6}
\begin{array}{l}
\D\frac{1}{\pi N_0}\left(g_{02}/\Delta\right)_{\Delta=0}=
\left(c^2_i+1\right)\left[c^2_i+1-\Gamma\left<\frac{1}
{\tilde{\omega}_0}\right>\right]^{-1}\\
\\
\D\times\left[-i\left<\frac{e}
{\tilde{\omega}_0}\right>
+\Gamma\left(c^2_a+1\right)^{-1}
\left<\frac{f^2}{\tilde{\omega}_0}\right>
\frac{1}{\pi N_0}\left(g_{22}/\Delta\right)_{\Delta=0}\right]
\end{array}
\end{equation}

\noindent
and 

\begin{equation}
\label{eIII7}
\begin{array}{l}
\D\frac{1}{\pi N_0}\left(g_{22}/\Delta\right)_{\Delta=0}= 
-i\left(c^2_a+1\right)\\
\\
\D\times\left[\left(c^2_a+1-\Gamma\left<\frac{f^4}
{\tilde{\omega}_0}\right>\right)\left(c^2_i+1-\Gamma\left<\frac{1}
{\tilde{\omega}_0}\right>\right)-\Gamma^2\left<\frac{f^2}
{\tilde{\omega}_0}\right>^2\right]^{-1}\\
\\
\D\times\left[\left<\frac{ef^2}
{\tilde{\omega}_0}\right>\left(c^2_i+1-\Gamma\left<\frac{1}
{\tilde{\omega}_0}\right>\right)+\Gamma\left<\frac{f^2}
{\tilde{\omega}_0}\right>\left<\frac{e}{\tilde{\omega}_0}\right>\right]
\end{array}
\end{equation}

\noindent
with $\tilde{\omega}_0=\omega+i\Sigma_0\left({\bf k},\omega\right)_{\Delta=0}$.  
In the following subsections we consider 
the above self-energies in the Born and unitary scattering limits.   

\subsection{Born scattering}
When both isotropic and anisotropic impurity scattering channels are in 
the Born scattering regime i.e. $c_i\gg 1$ ($v_iN_0\ll 1$),   
$c_a\gg 1$ ($v_aN_0\ll 1$) only the lowest 
order terms in the impurity potential play role in the self-energies. Keeping up  
to the square terms in $v_i$ ($v_a$) we obtain from Eqs. (\ref{eIII2}) and 
(\ref{eIII4})

\begin{equation}
\label{eIII11}
\Sigma_0\left({\bf k},\omega\right)_{\Delta=0}=
-i\pi nN_0\left[v^2_i+v^2_af^2\left({\bf k}\right)\right]sgn\left(\omega\right)
\end{equation}

\begin{equation}
\label{eIII12}
\Sigma_3\left({\bf k},\omega\right)_{\Delta=0}=
n\left[v_i+v_af^2\left({\bf k}\right)\right]
\end{equation}

\noindent
Though $\Sigma_3$ is nonzero, it may be absorbed into the chemical 
potential and its effect vanishes in the Born scattering. \cite{II1}  
For  
$\left(\Sigma_2\left({\bf k},\omega\right)/\Delta\right)_{\Delta=0}$ we 
obtain 

\begin{equation}
\label{eIII13}
\left(\Sigma_2\left({\bf k},\omega\right)/\Delta\right)_{\Delta=0}=
i\pi nN_0\left[v^2_i\left<e\right>+v^2_a\left<ef^2\right>
f^2\left({\bf k}\right)\right]\frac{sgn\left(\omega\right)}{\omega}
\end{equation}

\noindent
It is noteworthy that apart from the average value of the order parameter 
symmetry function, $\left<e\right>$, a term reflecting the overlap between 
$e\left({\bf k}\right)$ and $f^2\left({\bf k}\right)$ influences the self-energy 
$\left(\Sigma_2\left({\bf k},\omega\right)/\Delta\right)_{\Delta=0}$.  
The limit of $v_a=0$ gives the standard $s$-wave impurity scattering in 
the Abrikosov-Gorkov approximation \cite{II1} with  
$\Sigma_0\left({\bf k},\omega\right)_{\Delta=0}=-i\pi nN_0v^2_isgn\left(\omega\right)$,  
$\left(\Sigma_2\left({\bf k},\omega\right)/\Delta\right)_{\Delta=0}=
i\pi nN_0v^2_i\left<e\right>sgn\left(\omega\right)/\omega$, and the critical temperature 
determined by $ln\left(T_c/T_{c_{0}}\right)=\left(\left<e\right>^2-1\right)
\left[\Psi\left(1/2+nN_0v^2_i/\left(2T_c\right)\right)-\Psi\left(1/2\right)\right]$.   

\subsection{Unitary scattering}
The limit of the resonant impurity scattering in both isotropic and 
anisotropic channels i.e. $c_i\rightarrow 0$, $c_a\rightarrow 0$  
in Eqs. (\ref{eIII2}) and (\ref{eIII4}) leads to   

\begin{equation}
\label{eIII14}
\Sigma_0\left({\bf k},\omega\right)_{\Delta=0}=
-i\Gamma\left[1+f^2\left({\bf k}\right)\right]sgn\left(\omega\right)
\end{equation}

\noindent
and $\Sigma_3\left({\bf k},\omega\right)_{\Delta=0}=0$ which is consistent with 
the assumption of particle-hole symmetry of the excitation spectrum. 
The unitarity limit in Eqs. (\ref{eIII5})-(\ref{eIII7}) gives   

\begin{equation}
\label{eIII16}
\begin{array}{l}
\D\left(\Sigma_2\left({\bf k},\omega\right)/\Delta\right)_{\Delta=0}=
i\Gamma\left\{\left[1-\Gamma\left<\frac{1}{\tilde{\omega}_0}\right>
\right]^{-1}\left<\frac{e}{\tilde{\omega}_0}\right>\right.\\
\\
\D+\left[\Gamma\left(1-\Gamma\left<\frac{1}{\tilde{\omega}_0}\right>\right)^{-1} 
\left<\frac{f^2}{\tilde{\omega}_0}\right>+f^2\left({\bf k}\right)\right]\\ 
\\
\D\times\left[\left(1-\Gamma\left<\frac{1}{\tilde{\omega}_0}\right>\right)
\left(1-\Gamma\left<\frac{f^4}{\tilde{\omega}_0}\right>\right)-\Gamma^2
\left<\frac{f^2}{\tilde{\omega}_0}\right>^2\right]^{-1}\\
\\
\D\left.\times\left[\left(1-\Gamma\left<\frac{1}{\tilde{\omega}_0}\right>\right)
\left<\frac{ef^2}{\tilde{\omega}_0}\right>+\Gamma
\left<\frac{f^2}{\tilde{\omega}_0}\right>
\left<\frac{e}{\tilde{\omega}_0}\right>\right]\right\}sgn\left(\omega\right)
\end{array}
\end{equation}

\noindent
The impurity potential may lead to the strong scattering in one channel 
and the weak scattering in another. If the strong scattering takes 
place in the isotropic channel and the scattering in the anisotropic 
one is in the Born limit, then we deal with a case of the isotropic 
unitary scattering. \cite{II2,II3,III1,III2,III3,III4} 
The opposite case   
with the strong scattering in the anisotropic channel and the weak 
scattering in the isotropic one is equivalent to a nonphysical situation 
of unitary scattering in the anisotropic channel alone 
as discussed in section II.

\section{Critical temperature}
The critical temperature in the weak-coupling BCS approximation is determined by the equation

\begin{equation}
\label{eIV1}
\begin{array}{l}
\D ln\left(\frac{T_c}{T_{c_{0}}}\right)=
\pi T_c\sum_{\omega}\left[\frac{1}{\pi N_0}\sum_{{\bf k}}e\left({\bf k}\right)
\frac{\D e\left({\bf k}\right)+\frac{1}{i}
\left(\Sigma_1\left({\bf k},\omega\right)/\Delta\right)_{\Delta=0}
+\frac{1}{i}
\left(\Sigma_2\left({\bf k},\omega\right)/\Delta\right)_{\Delta=0}}
{\D\left(\omega+i\Sigma_0\left({\bf k},\omega\right)_{\Delta=0}\right)^2
+\left(\xi_k+\Sigma_3\left({\bf k},\omega\right)_{\Delta=0}\right)^2}\right.\\
\\
\D\left.-\frac{sgn\left(\omega\right)}{\omega}\right]
\end{array}
\end{equation}

\noindent
where because of the momentum-dependent self-energies the summations  
over the quasimomentum vector ${\bf k}$ restricted to the Fermi surface 
and Matsubara frequency $\omega$ need  
to be performed numerically. We do the calculations in two different scattering 
regimes - Born and unitary for the $d$-wave order parameter. 
Finally the results for the $(d+s)$-wave superconductor are discussed briefly. 
In order to proceed further one has to choose a function describing the anisotropy 
of the impurity potential. We study the functions given by the subsequent 
harmonics as they represent an orthonormal and complete set 
in the interval [0, 2$\pi$] and can be used for the Fourier expansion of any 
regular function. Harmonics up to the 4th order are analyzed and the results 
are extended to higher-order harmonic functions. It is important to realize that     
the impurity potential of $p$-wave anisotropy shows a particular property. Given by 
$sin\phi$ or $cos\phi$ function it represents a small-angle scattering which 
is relatively unimportant in contributing to the resistivity of the normal state. 
Taken even together with the $s$-wave scattering channel it results 
within the linear response approximation to the distribution function in a not well 
defined integrals determining the dc conductivity for some electric field orientations. 
\cite{II7} In other words, in the Boltzmann equation based analysis of the transport 
properties the $p$-wave scattering amplitude couples to the ''$1-cos(\theta)$" term 
in the collision integral. Therefore, in a fit to the real systems   
the $p$-wave scattering can be considered only as 
one of the components in the anisotropic channel of the impurity potential 
coexisting with some other higher-order harmonics. 
In order to emphasize this different feature of the $p$-wave potential among 
the other basis functions we present the momentum-dependent scattering with 
various amounts of the $p$-wave potential in a separate figure.  

The anisotropic impurity scattering is compared to the isotropic one of the 
same scattering strength. It means that we discuss the effect of substituting a part 
of a given isotropic impurity potential $v_0$ with an anisotropic term. The amplitude of   
the replaced isotropic potential, $v_0$, and the amplitudes of the isotropic and 
anisotropic scattering channels in the studied potential are related through the 
formulas $v_i=\alpha v_0$, and $v_a=\left(1-\alpha\right)v_0$, where the coefficient 
$\alpha$ ($0\leq\alpha\leq 1$) defines the partition of the potential. 

\subsection{Born scattering}
The impurity scattering is analyzed as a function of the pair-breaking parameter 
$\Gamma'=\pi nN_0v^2_0$ ($v_0=v_i+v_a$) which includes the overall scattering strength 
that is, takes both isotropic and 
anisotropic channels into account. It is convenient to use the ratio 
$v_a/v_i=\left(1-\alpha\right)/\alpha$ to define a particular potential. 

\subsubsection{s-wave superconductor}
We note, before discussing the unconventional superconductivity, that for the 
isotropic s-wave superconductor, given by $e\left({\bf k}\right)=1$, the $T_c$ equation 
(\ref{eIV1}) along with Eqs. (\ref{eIII11}) and (\ref{eIII13}) do not lead to 
a change of the critical temperature in agreement with the Anderson's theorem. 
\cite{IV1} 

\subsubsection{d-wave superconductor}
The critical temperature for a $d_{x^2-y^2}$-wave superconductor that is for  
$e\left({\bf k}\right)\sim (k^2_x-k^2_y)$ is presented for a large range of 
$v_a/v_i$ ratio values in Fig. 1. 
The major common feature of these diagrams is a lower $T_c$ suppression 
by the anisotropic impurity scattering compared to the isotropic ($s$-wave) one.  
According to their effect on superconductivity the anisotropic potentials can 
be classified in three groups defined by the function $f\left({\bf k}\right)$:  
$cos2\phi$, $sin2\phi$, and higher-order harmonics. 
The first of them, determined by $f\left({\bf k}\right)\sim cos2\phi$, leads  
to the strongest $T_c$ suppression. In this case the directions of the maximum impurity 
scattering correspond to the maxima of the order parameter and most of the 
pair-breaking process takes place in this region. Therefore, the suppression of 
superconductivity is particularly strong. However, for the level of anisotropy
up to $\left(1-\alpha\right)\approx 0.5$ ($v_a/v_i\approx 1$) the depairing 
effect of impurities is reduced with increasing contribution of $cos2\phi$ 
scattering compared to the isotropic scattering. Further increase 
of the anisotropic part in the impurity potential up to a level 
$\left(1-\alpha\right)\approx 0.6$ ($v_a/v_i\approx 1.5$) almost does not 
change the pair-breaking effect. When the contribution of the anisotropy 
exceeds this level the impurity effect on $T_c$ is enhanced and finally     
for $\left(1-\alpha\right)$ of the order of 0.83 ($v_a/v_i=5$) 
becomes comparable to the one of the $s$-wave scatterers as it is shown in 
Fig. 1e. 
The out of phase scattering takes place for $f\left({\bf k}\right)\sim sin2\phi$. 
This function is a $45^{\circ}$ rotation in the xy plane   
of the order parameter function $e\left({\bf k}\right)\sim cos2\phi$,  
that is, the impurity potential maxima correspond to the superconducting gap 
nodes and vice versa. In this way the impurity pair-breaking effect is minimized. 
The suppression of the critical temperature is reduced by an increasing amount of 
the anisotropic scattering in the impurity potential. For large $v_a/v_i$, however,  
the shape of the suppression lines changes slightly and a small enhanced 
suppression can be observed for some impurity concentration (Fig. 1e). 
In general, the anisotropic impurity 
potential given by a function orthogonal to the order parameter is less 
pair-breaking than $f\left({\bf k}\right)\sim e\left({\bf k}\right)$. The $T_c$ suppression 
by the 3rd and 4th order harmonics is less than the one of $cos2\phi$ but it exceeds  
that of $sin2\phi$. It is also decreasing for increasing level of anisotropy in the 
impurity potential up to $\left(1-\alpha\right)\approx 0.5$, then the pair-breaking effect 
of impurities practically saturates. The curvature of the graphs changes at some 
points, however, and the normalized critical pair-breaking parameter, 
$\Gamma'/\left(2\pi T_{c_{0}}\right)$, at which $T_c\approx 0$  may increase 
in some cases.  
It is worth observing that the critical temperature is almost equally suppressed  
by the impurity potentials given by the 3rd and 4th order harmonics  
for $v_a$ values up to $v_a/v_i=1$. This is particularly true for small anisotropic 
scattering levels. When the contribution of the anisotropic channel in the impurity 
potential is larger, the differences are more pronounced. Nevertheless, even for the 
amplitudes ratio as high as $v_a/v_i=2$\\  
( $1-\alpha\approx 0.67$ ) the curves 
differ only near a zero value of the critical temperature and the normalized 
critical pair-breaking parameters, $\Gamma'/\left(2\pi T_{c_{0}}\right)$,  
of these four harmonics are within an interval of the order of magnitude of $10^{-2}$. 
For the sake of transparency we present $cos3\phi$ scattering effect only in Fig. 1d. 
Though the curves of different 
harmonics overlap for the most of the critical temperature range for the amplitude 
in the anisotropic scattering channel five times as much as the one in the 
isotropic channel, $v_a/v_i=5$, they split distinctly at a low temperature of about 
$0.2\;T_c$ (Fig. 1e). A study of the higher order harmonics up to the 10th order for 
the anisotropic scattering strength equal to the isotropic one ($v_a/v_i=1$) leads 
to almost the same result as the 3rd and 4th order harmonics. This approximately universal 
behavior for harmonics from the 3rd to the 10th order is shown in Fig. 2. 
Concluding we may say that to a good accuracy the anisotropy of the impurity potential  
given by the harmonics of the order higher than two yields an approximately universal 
$T_c$ suppression. 
It is also worth mentioning that except for the isotropic scattering 
the critical temperature goes to a zero value asymptotically. The asymptotic  
tails start at very low temperatures of the order of magnitude of $10^{-4}\;T_{c_{0}}$ 
(not seen at the figure scale)  
where the fluctuation effects become important and may destroy superconductivity. 

Remembering that this potential cannot be considered on its own without any higher-order 
admixture, we show the pair-breaking effect of the $p$-wave scattering in the 
anisotropic channel for different $v_a/v_i$ values in Fig. 3. Compared to the other 
harmonics (Figs. 1-2) it yields the lowest $T_c$ suppression. This fact can be explained 
by an effectively small-angle scattering of the $p$-wave potential. If it is given  
by $cos\phi$ function then mostly the quasiparticles with their momenta parallel to the 
x-axis are affected, and for the $sin\phi$ function representing impurity potential the  
electrons moving along the y-axis are being scattered. Thus we deal with a weak 
practically one-dimensional scattering in a two-dimensional space. The pair-breaking  
effect of both basis functions $cos\phi$ and $sin\phi$ is the same within the accuracy of 
the numerical calculations. Worth observing is also a feature of a reduced $T_c$ 
suppression with an increasing amount of $p$-wave scattering in the impurity potential. 
However, at large amplitudes in the anisotropic channel ($v_a/v_i\sim 5$) the 
initial suppression of the critical temperature can be enhanced. Characteristic is  
also a clear asymptotic decrease of the critical temperature to its zero value. 

Any impurity potential given by Eq. (\ref{eI1}) can be represented as a combination 
of the potentials based on single harmonics. Therefore, we expect that the  
pair-breaking effect will be given by an appropriate superposition of the effects 
discussed above.  

\subsubsection{(d+s)-wave superconductor}
As an example of a $(d_{x^2-y^2}+s)$-wave superconductor we consider the one with 
an $s$-wave admixture of $10\%$. This is the order of magnitude of the $s$-wave level 
which cannot be ruled out by the ARPES measurements in the cuprates. \cite{IV2,IV3} 
The normalized to unity order parameter is given then by 
$e\left({\bf k}\right)=\left(cos2\phi+s\right)/
\left<\left(cos2\phi+s\right)^2\right>^{1/2}$, where 
$s=\left(0.005/0.99\right)^{1/2}$ and its FS average, $\left<e\right>=0.1$, corresponds 
to $10\%$ of the $s$-wave fraction in the $(d_{x^2-y^2}+s)$-wave superconductor. 
The results for $v_a/v_i=1$ are shown in Fig. 4. Even a small amount of the $s$-wave 
component, which is not destroyed by the potential scattering, results in a robustness 
of $T_c$ compared to the pure $d$-wave superconductor (Fig. 1c) and its asymptotic 
reduction. A change in the symmetry of the order parameter also causes a split in the 
suppression of the critical temperature by the $p$-wave anisotropic channel scattering. 

\subsection{Unitary scattering}
The relation determining the partition of the scattering amplitudes between the 
isotropic and anisotropic channels, $v_i+v_af\left({\bf k}\right)f\left({\bf k'}\right)=
v_0\alpha+v_0\left(1-\alpha\right)f\left({\bf k}\right)f\left({\bf k'}\right)$, holds 
as long as these amplitudes are finite. In the case of the resonant scattering 
we make an approximation of $v_i$ and $v_a$ diverging to infinity. Therefore 
we cannot control the relative scattering strengths in both channels and these 
processes become independent. Each part of the impurity potential enters the self-energy 
functions (Eqs. \ref{eIII2}-\ref{eIII7}) through variables $c_i$ and $c_a$. 
It is important to note that the self-energies depend on $c_i$ and $c_a$ 
parameters separately so they are functions of $c_i$ and $c_a$ and not of any 
combination of them (like $c_a/c_i$ for instance). 
In the unitarity limit $c_i\rightarrow 0$, $c_a\rightarrow 0$ and the contributions from 
the isotropic and anisotropic channels are in fact equal. Similarly to the Born 
scattering limit, we want to discuss the effect of an anisotropic impurity potential 
that replaces the isotropic one. In order to do an appropriate comparison we need to 
refer to the isotropic scattering in two channels or equivalently to a regular isotropic 
scattering in one channel with a doubled impurity concentration $2n$. Therefore, to 
compare the anisotropic unitary scattering with a corresponding isotropic one, the  
impurity concentration for the anisotropic scattering model must be half of the 
concentration of the $s$-wave impurities. We take the impurity 
concentration equal to   
$n/2$ for the two-channel anisotropic potential and the results are plotted as a function 
of $\Gamma'=2\Gamma=n/\left(\pi N_0\right)$, where $n$ is the real impurity concentration 
the same as in the Abrikosov-Gorkov scaling function for the $s$-wave scattering. That is,  
the two-channel scattering potential is averaged over two impurities so there is only 
one scattering channel per impurity present. This procedure introduces the anisotropic 
scattering potential of the form given by Eq. (\ref{eI1}) and the same scattering strength 
as the isotropic one in the unitarity limit.

\subsubsection{s-wave superconductor}
The lack of influence of the potential anisotropic impurity scattering 
in the unitary limit on the critical temperature can be shown rigorously for a small
concentration of the defects $n/(\pi N_0)\ll 1$, when the self-energies 
$\Sigma_0\left({\bf k}\right)_{\Delta=0}=-i\Gamma\left[1+f^2\left({\bf k}\right)\right]
sgn\left(\omega\right)$ (Eq. (\ref{eIII14})) and $\left(\Sigma_2\left({\bf k}\right)
/\Delta\right)_{\Delta=0}=-\Sigma_0\left({\bf k}\right)_{\Delta=0}/\omega$ 
(Eq. (\ref{eIII16})). For a larger impurity concentration a numerical analysis confirms 
this result with a very good accuracy. 

\subsubsection{d-wave superconductor}
The effect of the anisotropic unitary scattering on a $d_{x^2-y^2}$-wave superconductor 
is shown in Fig. 5. The same general rule as for the Born scattering holds here. 
The weakest pair-breaking effect is caused by the impurity potential involving $sin2\phi$ 
and most suppression is seen for the potential containing $cos2\phi$, that is the 
anisotropic scattering channel in phase with the order parameter. All the other harmonics 
orthogonal to $cos2\phi$ yield a moderate suppression of the critical temperature 
comparable to the isotropic scattering. They also form an almost universal suppression 
curve. A study of the $T_c$ reduction by the harmonics of the order up to ten  
shows the same approximate universality as that in Fig. 2, with the 
differences in the normalized critical pair-breaking parameter, $\Gamma'/ 
\left(2\pi T_{c_{0}}\right)$, of the order of $10^{-3}$. The curves of the 3rd and 4th 
order harmonics overlap in Fig. 5. Particularly interesting feature of the resonant 
scattering is a strong pair-breaking effect of the $cos2\phi$ anisotropic potential. 
It destroys superconductivity even faster than the isotropic impurities.  
Another characteristic fact is practically the same critical temperature dependence 
on the impurity concentration (i.e. not distinguishable in Fig. 5) for the $p$-wave 
scattering in the anisotropic channel and $sin2\phi$ scattering potential. 
Similarly to the weak scattering limit, 
except for the $s$-wave scattering the suppression of superconductivity at low $T_c$ 
is asymptotic, however not seen at the scale of Fig. 5.  

\subsubsection{(d+s)-wave superconductor}
A $(d_{x^2-y^2}+s)$-wave superconductor is more robust against 
the impurity scattering due to a nonzero $s$-wave component. Since $cos2\phi$ function is 
no longer in phase with the order parameter its pair-breaking effect is lowered 
and becomes even less than the isotropic one for low critical temperatures. 
The results for the same level of the $s$-wave part in a $(d_{x^2-y^2}+s)$-wave 
order parameter as the one discussed in the Born limit are shown in Fig. 6. 
It is worth mentioning that the $p$-wave scattering leads again to the same 
$T_c$ suppression as the $sin2\phi$ anisotropic potential. 

\section{Comparison to experiment}
In the overdoped samples $T_c/T_{c_{0}}$ data points plotted vs. impurity 
concentration form a universal curve independent of the critical temperature 
in the absence of impurities. \cite{I10,V0} Such a universal scaling behavior is 
characteristic of the impurity limited superconductivity provided the pair-breaking 
parameter $\Gamma'/\left(2\pi T_{c_{0}}\right)$ does not change with $T_{c_{0}}$. 
This requirement is equivalent to a constraint $N_0 T_{c_{0}}=constant$ in the unitarity 
limit and may be obeyed in the overdoped systems where the critical temperature 
decreases with increasing hole concentration. \cite{I10} There are also indications 
that a large residual resistivity due to Zn atoms in the cuprates corresponds to 
an impurity potential scattering in the unitary limit. \cite{I2} 
Thus, we consider the case of the unitary scattering focusing on the effect of Zn dopant.  
Working in the Born scattering limit requires an estimation of 
the impurity scattering potential which can be obtained from the residual 
resistivity in the normal state. \cite{I20,I21,I22,II7,I24} In that analysis  
a clear distinction between the impurity scattering life time and the 
transport relaxation time, which may differ if the impurity potential is 
anisotropic, \cite{II7} is needed.  
In order to compare our unitary scattering results with the experiment 
we have to convert the pair-breaking 
parameter $\Gamma'/\left(2\pi T_{c_{0}}\right)$ into the impurity concentration. 
As $\Gamma'=n/\left(\pi N_0\right)$ only the values of the density of states on 
the Fermi surface and the critical temperature in the absence of impurities are needed 
for that purpose. The density of states at the Fermi level is estimated from the  
measurements of the specific heat jump at the phase transition $\Delta C$. \cite{I1,V1}  
We employ the BCS weak-coupling relation $\Delta C/\gamma T_c\approx 1.43$ to obtain 
the normal-state Sommerfeld constant $\gamma$ which gives the density 
of states through $\gamma =2\pi^2k_B^2N_0/3$. 
It is important to note that the strong-coupling corrections \cite{V2,V3} as well as  
the interaction with impurities \cite{V4,V5} may change this 
relation significantly and in consequence alter the overall density of 
states. It may result in a wide range of $T_c$ solutions. \cite{V6} 
Because of a difficulty in the separation of lattice and electron contributions to the 
specific heat the thermodynamic experiments provide $\Delta C$ values with the accuracy  
depending on the quality of a sample.  
For high purity, fully oxygenated $Y\!-\!123$ compound grown in $BaZrO_3$ 
the mean-field  component of the electronic specific heat jump is estimated as 
$56 \pm 2 \left(mJ/K^2 mole\right)$. \cite{V1,V7,V8} In $La\!-\!214$ system this 
quantity is in the range of $14\pm 5 \left(mJ/K^2 mole\right)$. \cite{V1,V9}  
There is no specific heat jump observed at the phase transition in $Bi\!-\!2122$ 
compound \cite{V1} thus the present method of obtaining the density of states cannot be 
applied in this case. We discuss the experimental results in this system together 
with $Y\!-\!123$ and $La\!-\!214$ compounds more extensively in Ref. 53.  
The density of states calculated from the above $\Delta C$ values are used for 
the evaluation of the impurity concentration, which is represented as the per cent  
number of impurities (defects) per planar Cu site. We do the calculations with 
a fixed density of states as we do not have any quantitative data showing its 
change with doping or disorder. Therefore, the results are not really universal  
and depend on the critical  
temperature in the absence of impurities $T_{c_{0}}$. We group them according to  
the values of the critical temperature of a pure system.  
In the optimally doped $Y\!-\!123$ compound $T_{c_{0}}$ varies in a very 
narrow range of values and for the theoretical calculation we take $T_{c_{0}}=91K$. 
The phase diagrams for the $d$-wave superconductor together with the experimental 
data of Zn doped samples \cite{I13,I14,I15,I16,I17} are shown in Fig. 7. 
The scattering effect of a given potential corresponds to the area between two curves 
of the same style, for instance the solid lines stand for the $sin2\phi$ impurity potential. 
This broadened range of values stem from the uncertainty in the estimation of the 
FS density of states, that is, from the accuracy of the $\Delta C$ values.  
We note, that all the experimental points fall in 
the region of the theoretically predicted $T_c$ suppression by the unitary  
impurity scattering. Interestingly, the pair-breaking effect of Zn atoms 
in certain samples corresponds to the resonant scattering in different 
anisotropic channel of the impurity potential. We can clearly distinguish 
data that can be approximated by the scattering potentials given by $cos2\phi$ 
and $sin2\phi$ functions. The rest of the experimental points lie in the range 
of isotropic scattering or anisotropic scattering with higher order harmonics.  
Although the critical temperatures of the analyzed samples are very close, we do not see 
a universal suppression dependence as in $Bi\!-\!2122$ and $La\!-\!214$ compounds. 
\cite{I10,V0} This feature may suggest a possible sample-dependence of the effective impurity 
scattering caused for instance by the differences in a sample preparation or 
Zn substitution processes. Even small changes in the hole concentration alter the 
electronic density of states at the Fermi level and result in a modified impurity 
scattering rate.  

According to $T_{c_{0}}$ values, we gather the experimental 
data for Zn doped $La\!-\!214$ in two groups corresponding to $T_{c_{0}}$ equal to 
$30K$ (Fig. 8a) \cite{I1} and $36K$ (Fig. 8b). \cite{I1,I3} 
For the sample of $T_{c_{0}}\approx 30K$ the experimental  
suppression of the critical temperature is in the range of anisotropic scattering 
determined by $sin2\phi$ (Fig. 8a). The pair-breaking effect of Zn atoms in the 
samples characterized by $T_{c_{0}}\approx 36K$ is on the edge of anisotropic 
scattering in higher order harmonics but also right in the middle of $sin2\phi$ 
based anisotropic scattering. As we can see, the quantitative calculations for  
$La\!-\!214$ compound contain a large uncertainty margin which is caused by a 
lack of a precise value of the specific heat jump at the phase transition and 
consequently of the electron density of states on the Fermi surface. The error in 
the experimentally estimated magnitude of $\Delta C$ is of the order of $36\%$.      

A similar analysis of Ni substituted samples shows that the pair-breaking effect in the 
unitary limit is stronger than observed experimentally. For the sake of comparison 
we present Ni doped $Y\!-\!123$ compound data \cite{I12,I14,I15,I18} and our theoretical 
curves for the resonant scattering in Fig. 9. Less difference between the 
experimental and theoretical results is seen for $La\!-\!214$. \cite{V6}  
Although this comparison is suggestive for a weak potential scattering, the 
detailed calculations in the Born scattering limit giving the critical temperature 
dependence on the residual resistivity \cite{II7} are needed in order to draw more 
firm conclusions.  

Finally we discuss the electron irradiation experiments in $Y\!-\!123$. 
\cite{I19,I20,I21,I21a} The results for low-energy (60-120 keV) incident electrons 
read from Fig. 12 of Ref. 21 are shown in Fig. 10. The experimental points are in the 
range of the pair-breaking effect of the scattering potential with the anisotropy given 
by $sin2\phi$. The initial $T_c$ suppression, however, is more gradual than the 
one obtained from the theoretical calculation and cannot be explained by the standard 
impurity pair-breaking mechanism.  
Unfortunately, we do not have any other set of electron irradiation data representing 
the change of the critical temperature with defect concentration. The analysis 
by Tolpygo et. al. \cite{I21} shows that the initial $T_c$ suppression for low-energy 
(100 keV) electron irradiation reported by Legris et al. \cite{I21a} agrees with the one 
shown in Fig. 10. For higher-energy electrons \cite{I19} $T_c$ is reduced initially about 
twice as much as in Ref. 21 which becomes in the range of the $sin2\phi$ determined  
behavior for low defect concentration. 

\section{Conclusions}
We have studied the pair-breaking effect of the anisotropic impurity 
scattering in the t-matrix approximation for 
$d_{x^2-y^2}$-wave and $(d_{x^2-y^2}+s)$-wave superconductors.  
The Born and the unitary limits have been discussed analytically and numerically.  
Although limited to the employed phenomenological model the conclusions we draw should be 
suggestive to a larger class of potentials capturing the feature of anisotropy. 

We have shown, that the effect of the anisotropic impurity scattering  
can be considered in four groups for the Born scattering and three groups in 
the case of unitary scattering which are determined by the form of the function 
$f\left({\bf k}\right)$ defining symmetry of the scattering potential. 
In both scattering limits the strongest suppression of 
the critical temperature is caused by the impurity potential given by 
$f\left({\bf k}\right)\sim cos2\phi$ which is in phase with the order parameter 
for the $d_{x^2-y^2}$ state and overlaps significantly with the $(d_{x^2-y^2}+s)$-wave 
state of a major $d$-wave component. The pair-breaking effect of this potential 
with a large amount of anisotropic amplitude can be comparable with the one   
of the isotropic scattering in the Born limit and exceeds the $s$-wave impurity effect 
in the unitary limit. This issue is particularly important as it shows that   
a weak reduction of $T_c$ due to anisotropy of the impurity potential proportional to the   
superconducting order parameter which follows from the weak-scattering     
model \cite{I24} is not a general feature of the anisotropic impurity scattering. 
Another class is defined by $f\left({\bf k}\right)\sim sin2\phi$. 
It leads to the lowest impurity pair-breaking effect in the case of the unitary 
scattering and second lowest for the Born limit. This kind of scattering 
is maximal in the direction of the nodes of the order parameter and it vanishes 
where the gap function has its maxima. Therefore, the effective scattering is minimized 
by the symmetry of the impurity potential. Any other function orthogonal to the 
$d$-wave order parameter results in a rather universal $T_c$ suppression and falls 
into the third group of the potentials. The pair-breaking in this case is less than 
that of the isotropic scattering but it exceeds the one of $sin2\phi$ based potential, 
and in the unitary limit is very close to the isotropic scattering. 
Resonant scatterers in the $p$-wave channel lead to the same $T_c$ 
suppression as the $d$-wave scattering given by $sin2\phi$ function. In the Born 
scattering limit, however, the $p$-wave anisotropic scattering results in the 
lowest $T_c$ suppression. 

We have compared our results for the $d_{x^2-y^2}$-wave superconductor with the experimental 
data assuming that the impurity potential is close to the unitary limit.  
Within the accuracy of our calculations the Zn atoms can be 
considered as the resonant scatterers in overdoped $La\!-\!214$ and $Y\!-\!123$  
compounds. The pair-breaking effect of the structure defects produced by the electron 
irradiation is also in the range of magnitude of the unitary scattering with the 
anisotropy of the impurity potential given by $sin2\phi$ function. The theoretical results 
strongly depend on the value of the electronic density of states on the Fermi surface. 
Thus far, for some systems (like $La\!-\!214$) there is only an estimation of this 
value with a large margin of uncertainty available.  
It opens a wide range of possible 
$T_c$ solutions and makes a proper analysis very difficult. Therefore, before any final 
conclusion about the impurity suppression of the critical temperature in the cuprates 
can be made, a much more accurate determination of the electronic density of states on the 
Fermi surface has to be done.  

\section*{Acknowledgments}
We would like to thank B. Nachumi for his help in analyzing the experimental data,  
as well as T. Kluge for the discussions on the effect of impurities in high-$T_c$ 
superconductors and pointing to us several important papers. We are also grateful to 
A. Junod for providing us with the specific heat data and for the enlightening 
explanations of the thermodynamic measurements in the cuprates. Finally, we 
would like to thank P. Hirschfeld for critical comments on the manuscript.  

This work was supported by the Natural Sciences and Engineering Research Council  
of Canada.

\newpage
\section*{Figure Captions}

\noindent
Fig. 1. Normalized critical temperature $T_c/T_{c_{0}}$ of the $d_{x^2-y^2}$-wave 
superconductor as a function of the normalized impurity scattering rate 
$\Gamma'/(2\pi T_{c_{0}})$ in the Born limit for the impurity potential symmetry 
(solid curves from the top): $sin2\phi$, $sin3\phi$ and higher-order harmonics, 
$cos2\phi$. The amount of anisotropy in the  potential is: 
(a) $\left(1-\alpha\right)\approx 0.17$ ($v_a/v_i=0.2$),
(b) $\left(1-\alpha\right)\approx 0.33$ ($v_a/v_i=0.5$), 
(c) $\left(1-\alpha\right)=0.5$ ($v_a/v_i=1.0$), 
(d) $\left(1-\alpha\right)\approx 0.67$ ($v_a/v_i=2.0$), 
(e) $\left(1-\alpha\right)\approx 0.83$ ($v_a/v_i=5.0$) the higher-order 
harmonics split in this case into (from the top): $sin4\phi$, $cos3\phi$, 
$sin3\phi$, $cos4\phi$. The dashed line represents isotropic suppression 
of the critical temperature.\\  

\noindent
Fig. 2. Shaded area represents the effect of the anisotropic impurity potentials given 
by $cosl\phi$ and $sinl\phi$ functions for $3\le l \le 10$ and $\left(1-\alpha\right)=0.5$  
($v_a/v_i=1.0$) in the $d_{x^2-y^2}$-wave state in the Born limit.\\  

\noindent
Fig. 3. The $d_{x^2-y^2}$-wave state $T_c$ suppression by the Born scattering 
$p$-wave impurity 
potential with $f\left({\bf k}\right)=sin\phi$ (or $cos\phi$) and the anisotropic 
scattering strength $\left(1-\alpha\right)$ ($v_a/v_i$) equal to (from the top) 
0.83 (5.0), 0.67 (2.0), 0.50 (1.0), 0.33 (0.5), 0.17 (0.2). The isotropic scattering 
is shown with the dashed line.\\ 

\noindent
Fig. 4. Normalized critical temperature $T_c/T_{c_{0}}$ for the $(d_{x^2-y^2}+s)$-wave 
superconductor ($\left<e\right>=0.1$) as a function of the normalized impurity 
scattering rate $\Gamma'/(2\pi T_{c_{0}})$ in the Born limit for the impurity potential 
symmetry (solid curves from the top): $cos\phi$, $sin\phi$, $sin2\phi$,   
$sin3\phi$ and higher-order harmonics, $cos2\phi$. The amount of anisotropy in the  
potential is $\left(1-\alpha\right)=0.5$ ($v_a/v_i=1.0$). The dashed curve represents 
the effect of the isotropic impurity scattering.\\

\noindent
Fig. 5. Normalized critical temperature $T_c/T_{c_{0}}$ of the $d_{x^2-y^2}$-wave 
superconductor as a function of the normalized impurity scattering rate 
$\Gamma'/(2\pi T_{c_{0}})$ in the unitary limit for the impurity potential 
symmetry (solid curves from the top): $sin2\phi$ (and $p$-wave), $sin3\phi$ 
and higher-order harmonics, $cos2\phi$. The dashed curve represents 
the effect of the isotropic impurity scattering.\\

\noindent
Fig. 6. Normalized critical temperature $T_c/T_{c_{0}}$ of the $(d_{x^2-y^2}+s)$-wave 
superconductor ($\left<e\right>=0.1$) as a function of the normalized impurity 
scattering rate $\Gamma'/(2\pi T_{c_{0}})$ in the unitary limit for the impurity potential 
symmetry (solid curves from the top): $sin2\phi$ (and $p$-wave), $sin3\phi$
and higher-order harmonics, $cos2\phi$. The dashed curve represents
the effect of the isotropic impurity scattering.\\ 

\noindent
Fig. 7. Critical temperature of Zn substituted optimally doped $Y\!-\!123$ samples 
vs impurity 
concentration per cent per planar Cu site, $n_{Zn}$. The data are taken from 
Refs. 13 (filled circles), 14 (open squares), 15 (diamonds), 16 (open circles), 
17 (filled squares). The area between the curves of the same sort corresponds 
to the pair-breaking effect of the unitary scattering impurities of the following 
anisotropies: $sin2\phi$  (solid), $cos3\phi$ and higher-order (long dashed), 
$cos2\phi$ (dashed). The isotropic impurity scattering is represented by the dot-dashed 
curves.\\

\noindent
Fig. 8. Critical temperature of Zn substituted $La\!-\!214$ samples vs impurity
concentration per cent per planar Cu site, $n_{Zn}$. The data are taken from 
Refs. 1 (filled squares), 3 (open squares). The theoretical curves in the unitary limit  
are drawn for two critical temperatures: (a) $T_{c_{0}}=30K$, (b) $T_{c_{0}}=36K$. 
The impurity potentials represented by different curves agree with the notation 
in Fig. 7.\\   

\noindent
Fig. 9.  Critical temperature of Ni substituted optimally doped $Y\!-\!123$ samples 
vs impurity concentration per cent per planar Cu site, $n_{Ni}$. 
The data are taken from Refs. 12 (open circles), 14 (filled squares), 15 (diamonds),
18 (open squares). The theoretically calculated unitary scattering curves correspond 
to the potentials described as in Fig. 7.\\   

\noindent
Fig. 10. Critical temperature of optimally doped $Y\!-\!123$ sample with in-plane 
oxygen vacancies induced by $60-120$ keV electron irradiation (Ref. 21) vs 
defect concentration per cent per planar Cu site, $n_{(O\;defects)}$. 
The theoretically calculated unitary scattering curves correspond to the 
potentials described as in Fig. 7.\\ 

\newpage
\begin{center}
\begin{figure}[p]
\parbox{0.1cm}{\LARGE\vfill $$T_c/T_{c_{0}}$$\vspace{1.5ex}\vfill }
\parbox{15cm}{\epsfig{file=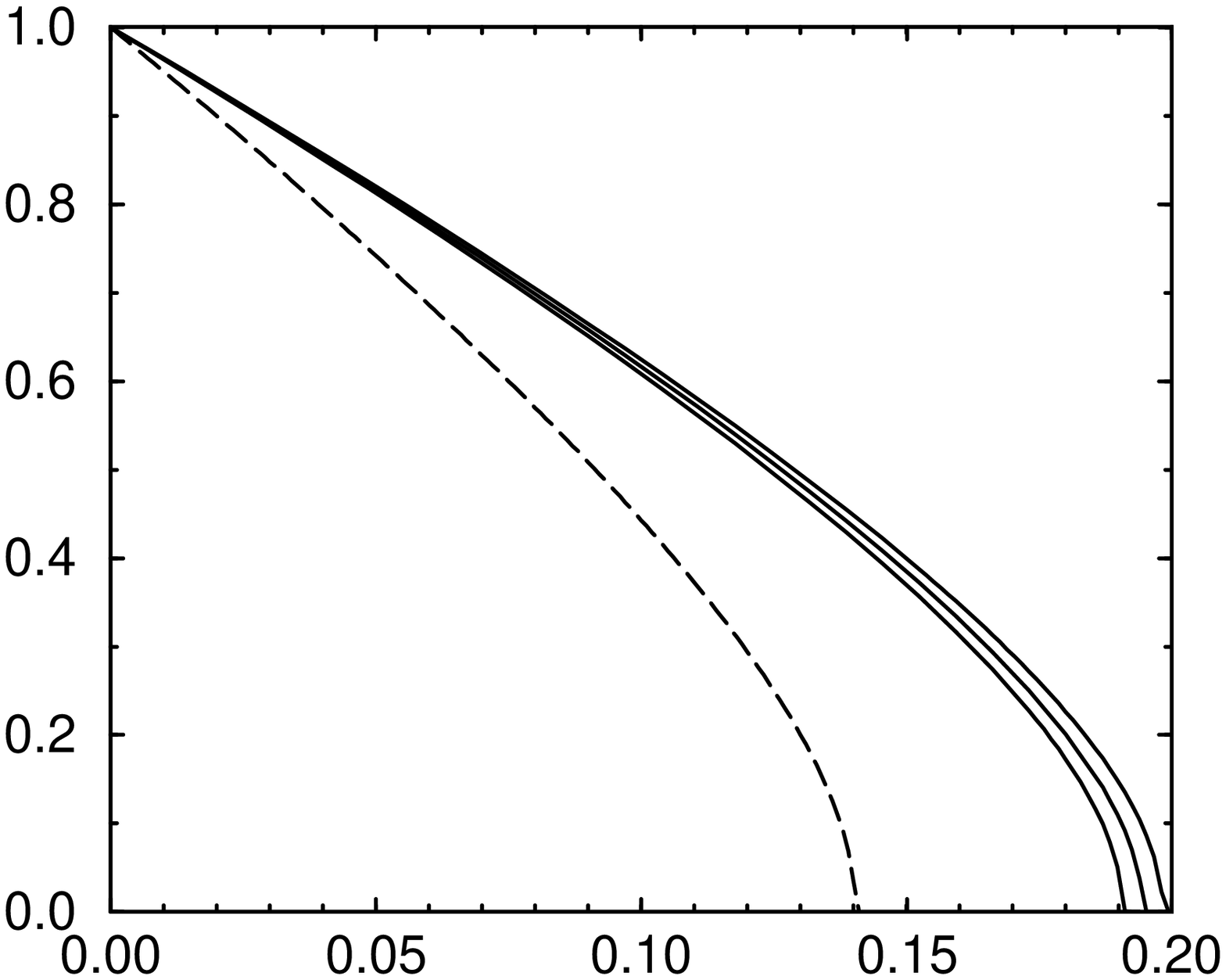,height=15cm,width=15cm} }
\parbox{0.5cm}{\hfill}
\parbox{18cm}{\LARGE\vspace{-6ex}\hfill $$\Gamma'/(2\pi T_{c_{0}})
\;\;\;\;\;$$\hfill }
\caption{Fig.1a:} 
\end{figure}
\end{center}

\newpage
\begin{center}
\begin{figure}[p]
\parbox{0.1cm}{\LARGE\vfill $$T_c/T_{c_{0}}$$\vspace{1.5ex}\vfill }
\parbox{15cm}{\epsfig{file=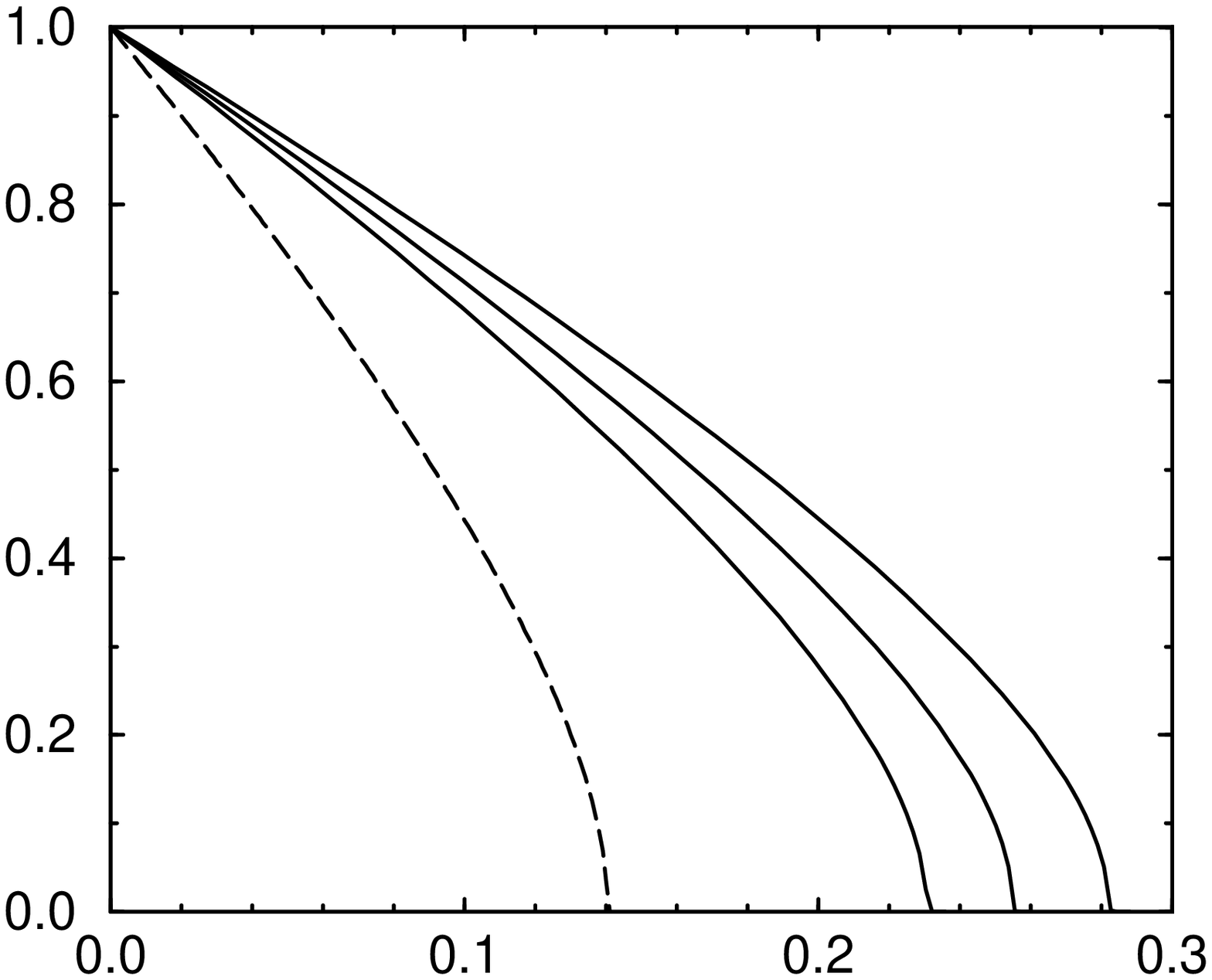,height=15cm,width=15cm} }
\parbox{0.5cm}{\hfill}
\parbox{18cm}{\LARGE\vspace{-6ex}\hfill $$\Gamma'/(2\pi T_{c_{0}})
\;\;\;\;\;$$\hfill }
\caption{Fig.1b:} 
\end{figure}
\end{center}

\newpage
\begin{center}
\begin{figure}[p]
\parbox{0.1cm}{\LARGE\vfill $$T_c/T_{c_{0}}$$\vspace{1.5ex}\vfill }
\parbox{15cm}{\epsfig{file=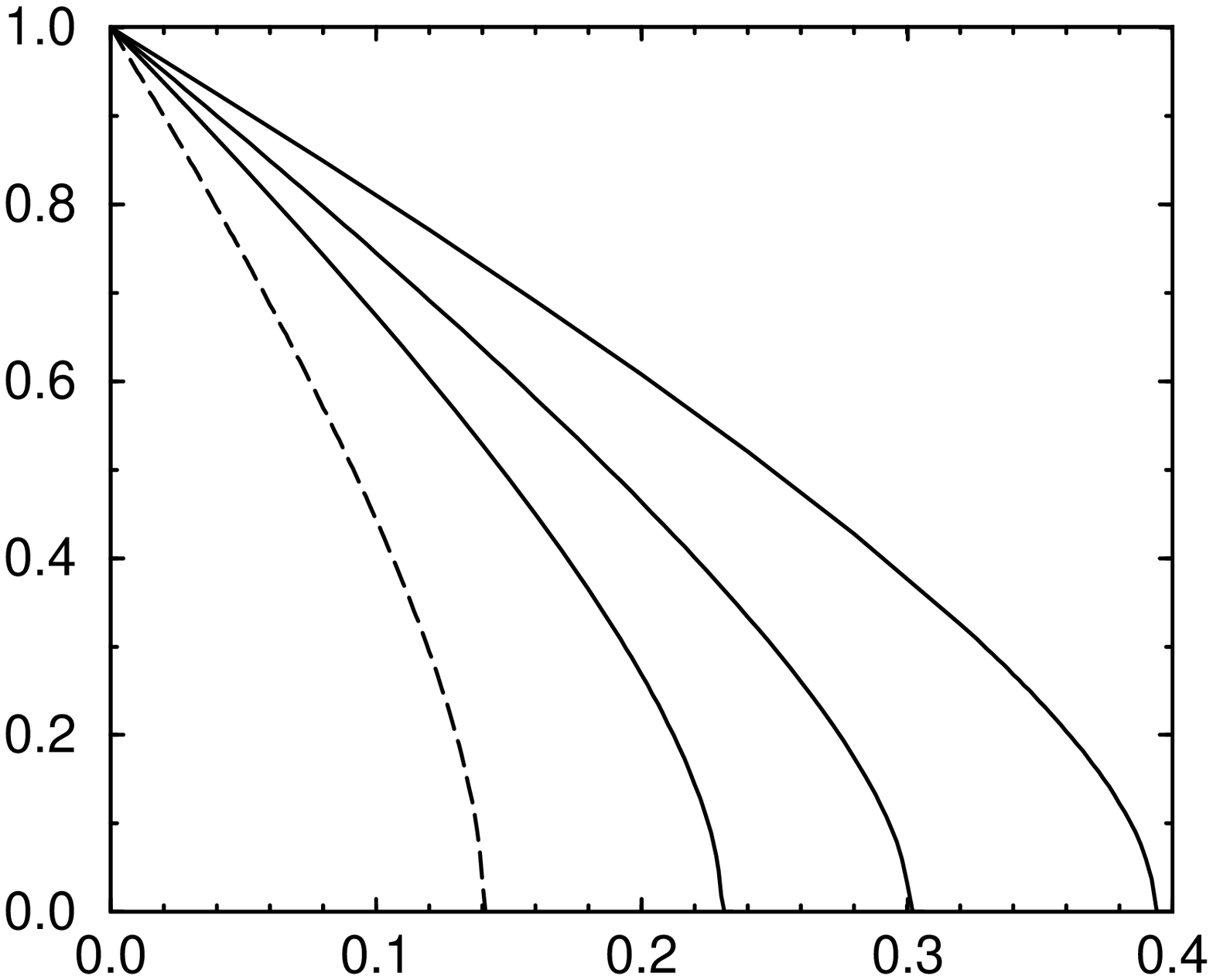,height=15cm,width=15cm} }
\parbox{0.5cm}{\hfill}
\parbox{18cm}{\LARGE\vspace{-6ex}\hfill $$\Gamma'/(2\pi T_{c_{0}})
\;\;\;\;\;$$\hfill }
\caption{Fig.1c:} 
\end{figure}
\end{center}

\newpage
\begin{center}
\begin{figure}[p]
\parbox{0.1cm}{\LARGE\vfill $$T_c/T_{c_{0}}$$\vspace{1.5ex}\vfill }
\parbox{15cm}{\epsfig{file=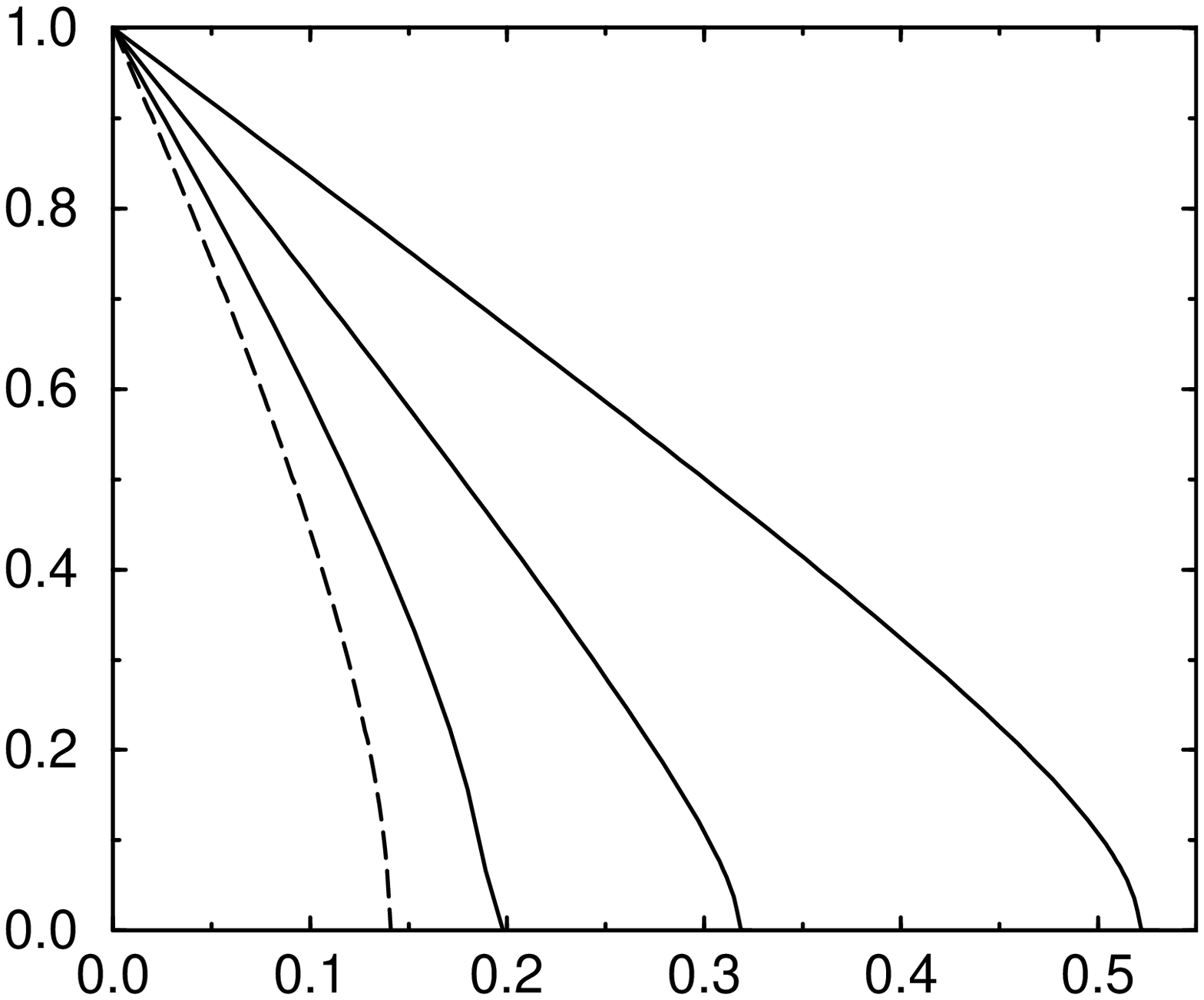,height=15cm,width=15cm} }
\parbox{0.5cm}{\hfill}
\parbox{18cm}{\LARGE\vspace{-6ex}\hfill $$\Gamma'/(2\pi T_{c_{0}})
\;\;\;\;\;$$\hfill }
\caption{Fig.1d:} 
\end{figure}
\end{center}

\newpage
\begin{center}
\begin{figure}[p]
\parbox{0.1cm}{\LARGE\vfill $$T_c/T_{c_{0}}$$\vspace{1.5ex}\vfill }
\parbox{15cm}{\epsfig{file=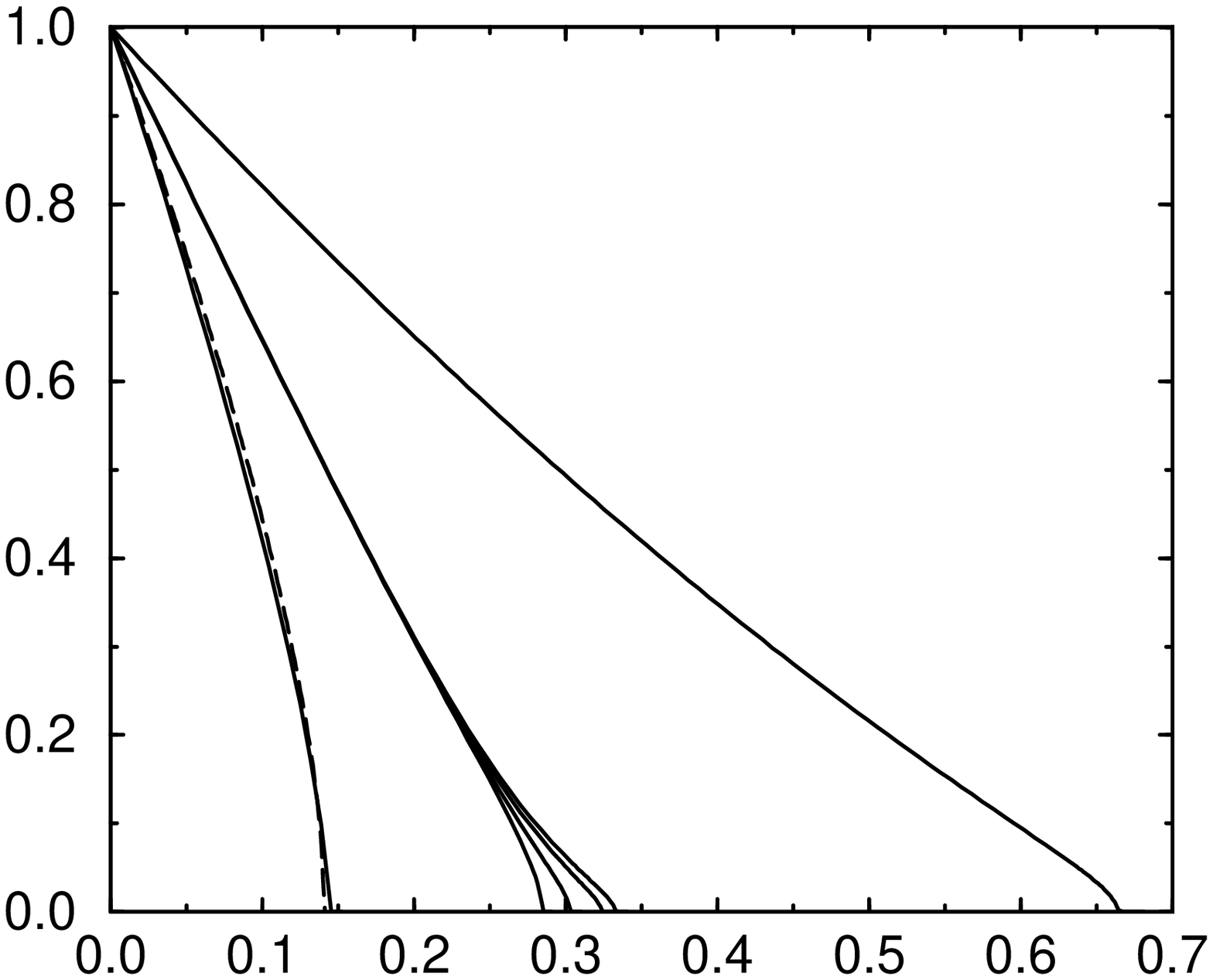,height=15cm,width=15cm} }
\parbox{0.5cm}{\hfill}
\parbox{18cm}{\LARGE\vspace{-6ex}\hfill $$\Gamma'/(2\pi T_{c_{0}})
\;\;\;\;\;$$\hfill }
\caption{Fig.1e:} 
\end{figure}
\end{center}

\newpage
\begin{center}
\begin{figure}[p]
\parbox{0.1cm}{\LARGE\vfill $$T_c/T_{c_{0}}$$\vspace{1.5ex}\vfill }
\parbox{15cm}{\epsfig{file=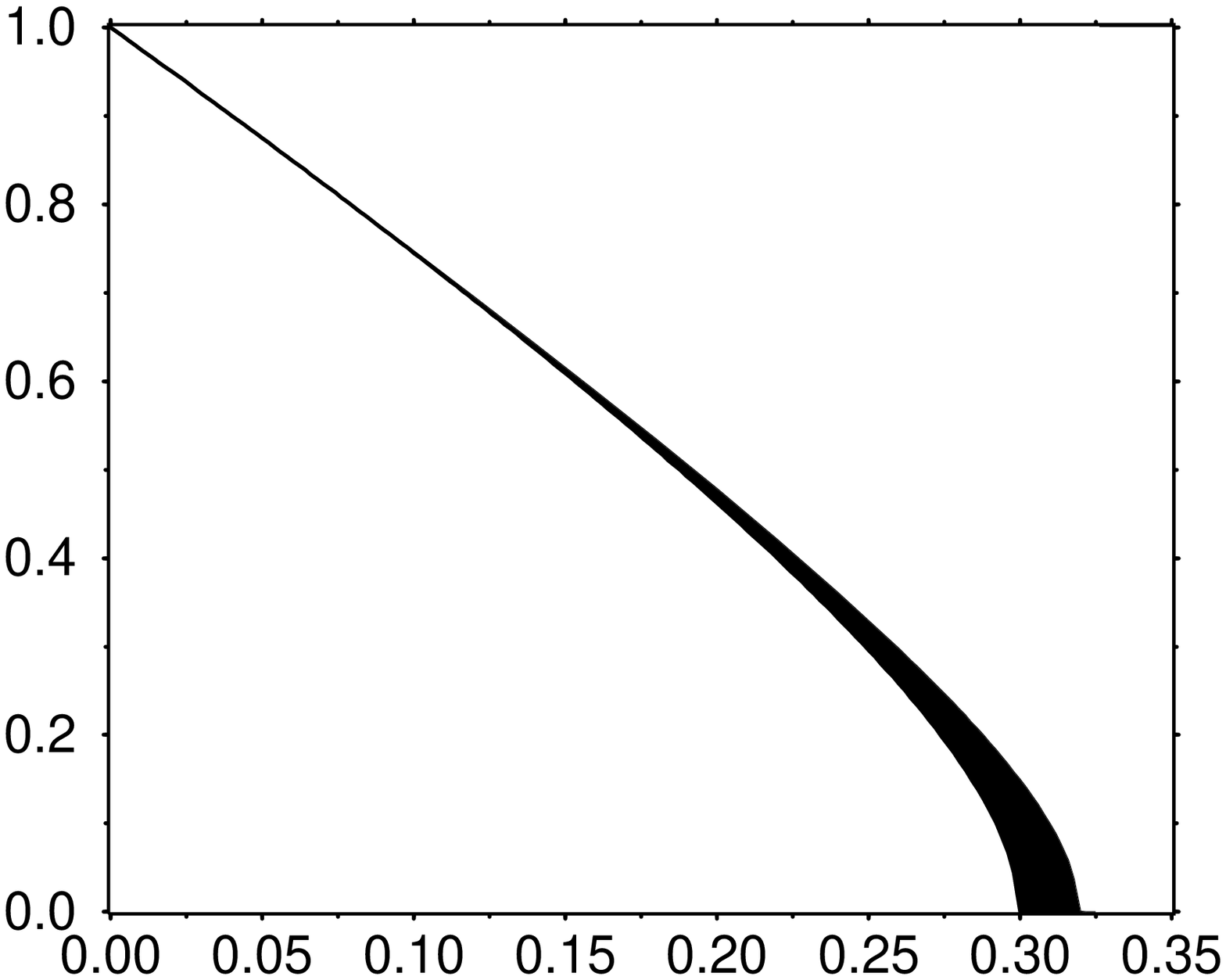,height=15cm,width=15cm} }
\parbox{0.5cm}{\hfill}
\parbox{18cm}{\LARGE\vspace{-6ex}\hfill $$\Gamma'/(2\pi T_{c_{0}})
\;\;\;\;\;$$\hfill }
\caption{Fig.2:} 
\end{figure}
\end{center}

\newpage
\begin{center}
\begin{figure}[p]
\parbox{0.1cm}{\LARGE\vfill $$T_c/T_{c_{0}}$$\vspace{1.5ex}\vfill }
\parbox{15cm}{\epsfig{file=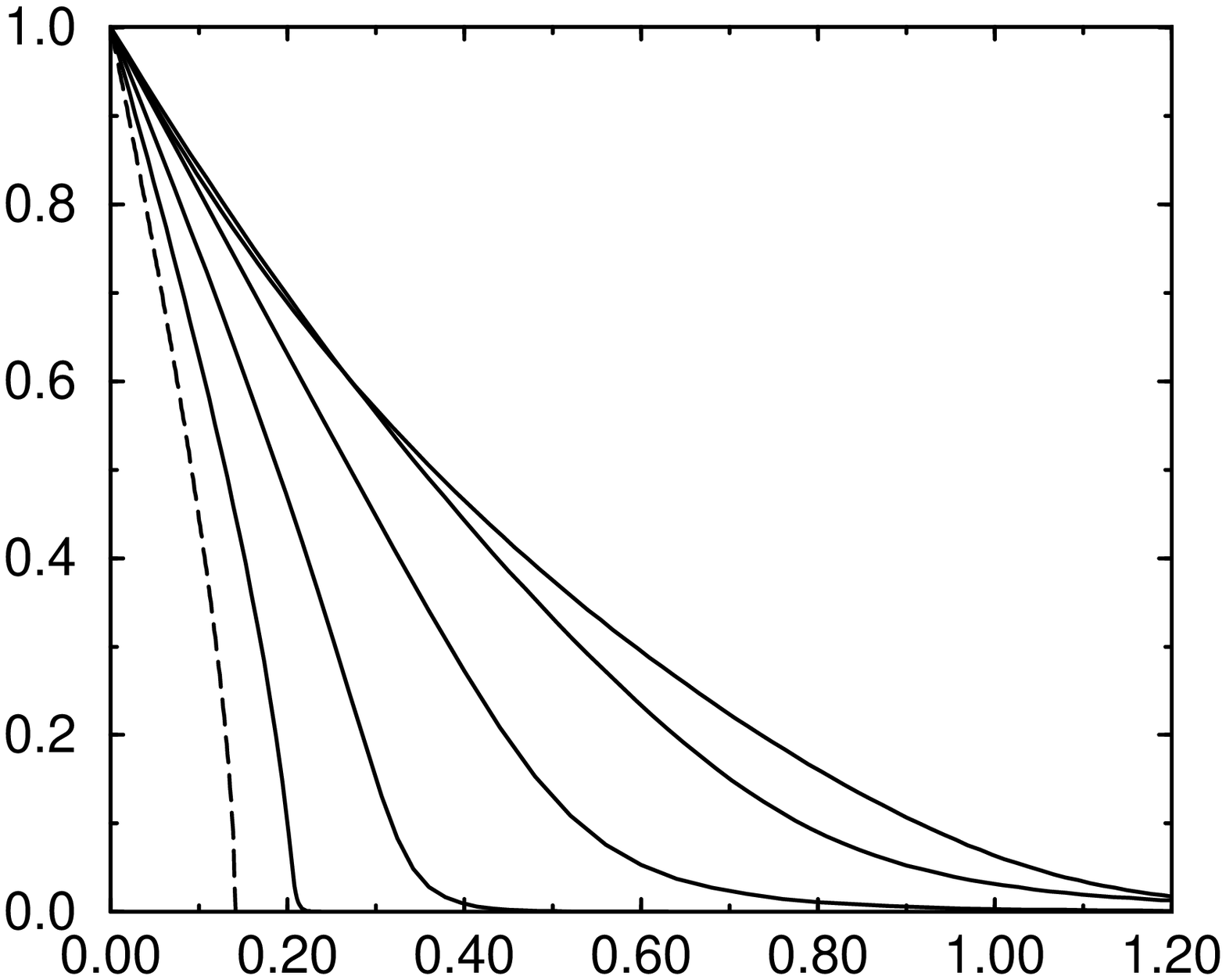,height=15cm,width=15cm} }
\parbox{0.5cm}{\hfill}
\parbox{18cm}{\LARGE\vspace{-6ex}\hfill $$\Gamma'/(2\pi T_{c_{0}})
\;\;\;\;\;$$\hfill }
\caption{Fig.3:} 
\end{figure}
\end{center}

\newpage
\begin{center}
\begin{figure}[p]
\parbox{0.1cm}{\LARGE\vfill $$T_c/T_{c_{0}}$$\vspace{1.5ex}\vfill }
\parbox{15cm}{\epsfig{file=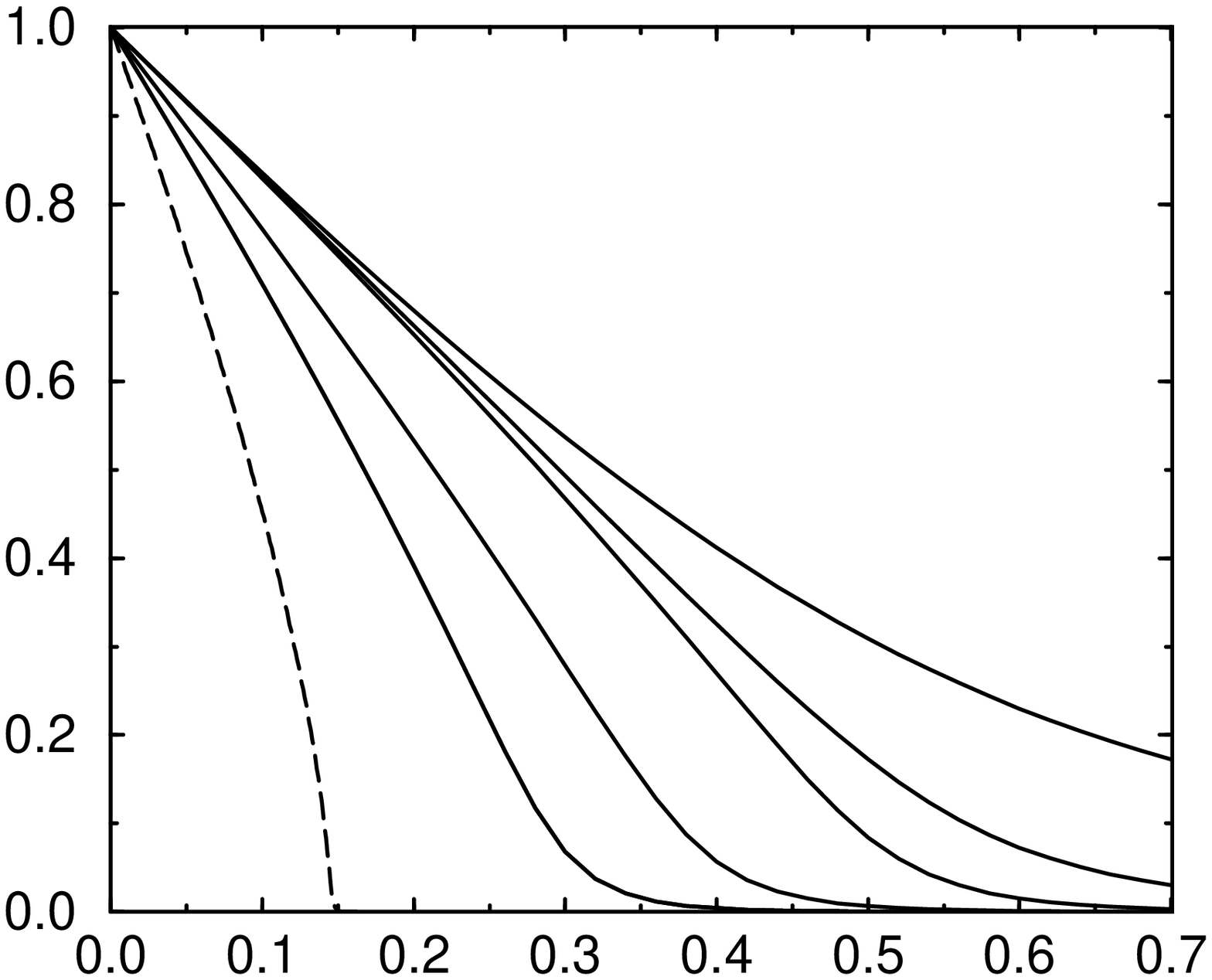,height=15cm,width=15cm} }
\parbox{0.5cm}{\hfill}
\parbox{18cm}{\LARGE\vspace{-6ex}\hfill $$\Gamma'/(2\pi T_{c_{0}})
\;\;\;\;\;$$\hfill }
\caption{Fig.4:} 
\end{figure}
\end{center}

\newpage
\begin{center}
\begin{figure}[p]
\parbox{0.1cm}{\LARGE\vfill $$T_c/T_{c_{0}}$$\vspace{1.5ex}\vfill }
\parbox{15cm}{\epsfig{file=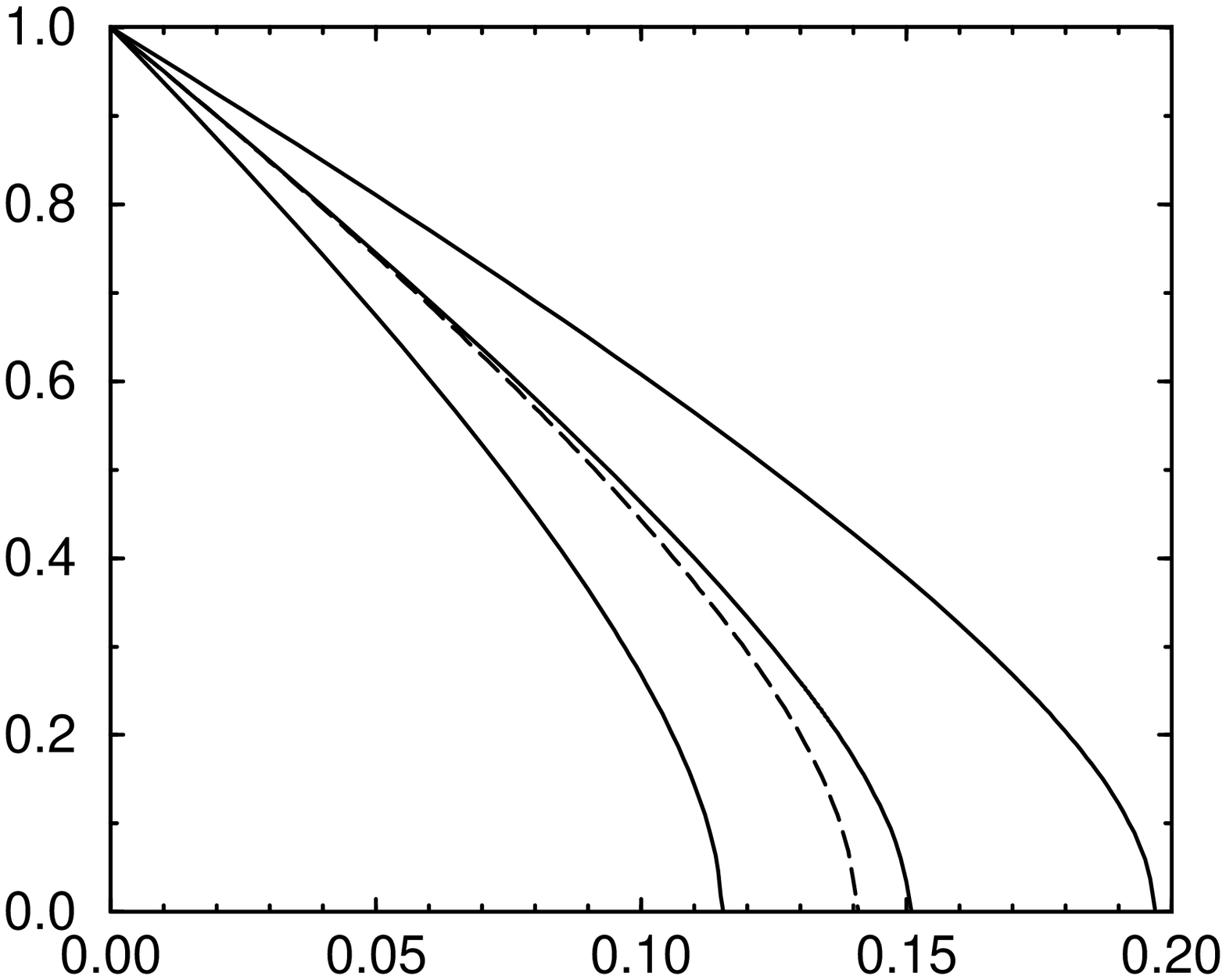,height=15cm,width=15cm} }
\parbox{0.5cm}{\hfill}
\parbox{18cm}{\LARGE\vspace{-6ex}\hfill $$\Gamma'/(2\pi T_{c_{0}})
\;\;\;\;\;$$\hfill }
\caption{Fig.5:} 
\end{figure}
\end{center}

\newpage
\begin{center}
\begin{figure}[p]
\parbox{0.1cm}{\LARGE\vfill $$T_c/T_{c_{0}}$$\vspace{1.5ex}\vfill }
\parbox{15cm}{\epsfig{file=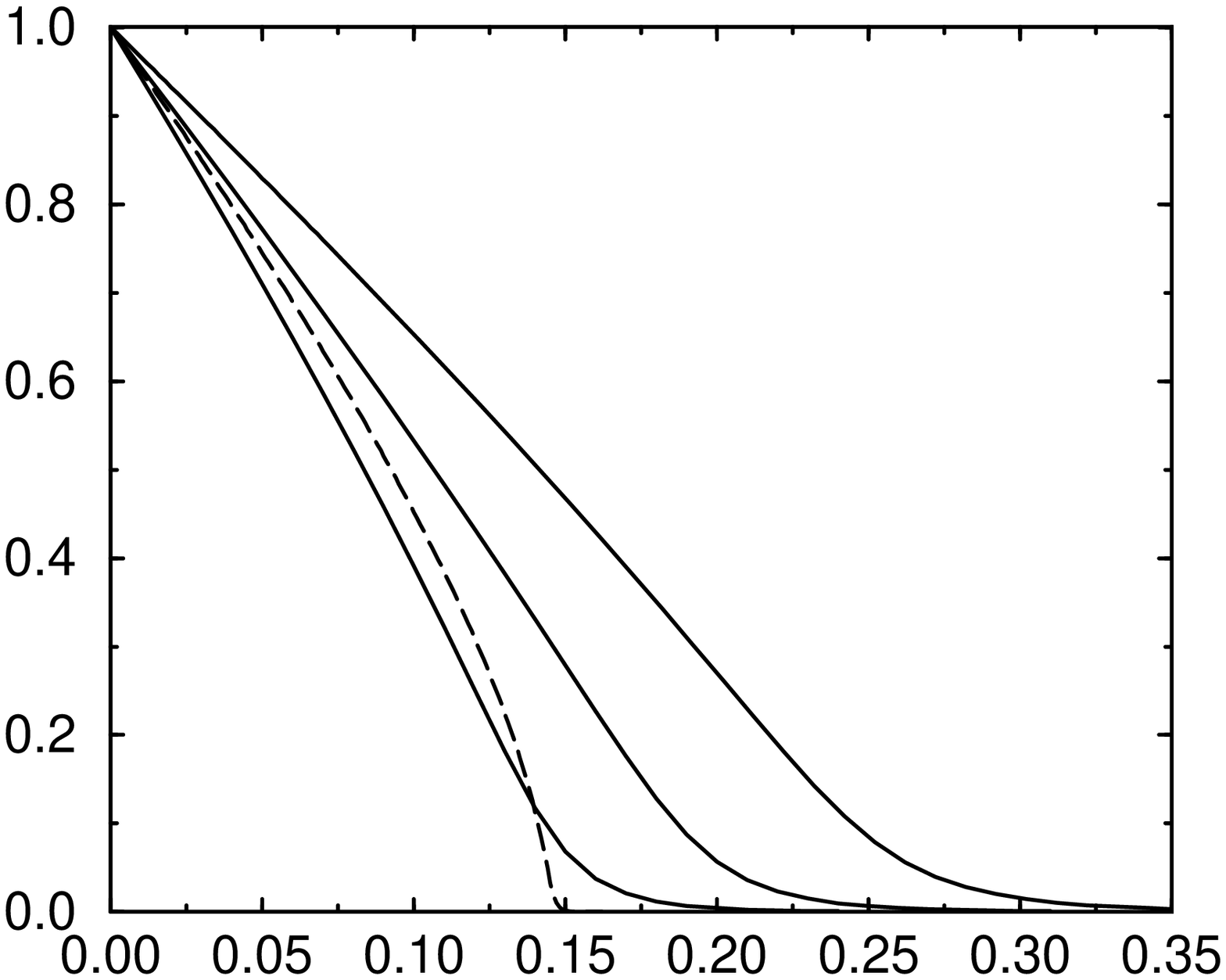,height=15cm,width=15cm} }
\parbox{0.5cm}{\hfill}
\parbox{18cm}{\LARGE\vspace{-6ex}\hfill $$\Gamma'/(2\pi T_{c_{0}})
\;\;\;\;\;$$\hfill }
\caption{Fig.6:} 
\end{figure}
\end{center}

\newpage
\begin{center}
\begin{figure}[p]
\parbox{0.1cm}{\LARGE\vfill $$T_c/T_{c_{0}}$$\vspace{0.5ex}\vfill }
\parbox{15cm}{\epsfig{file=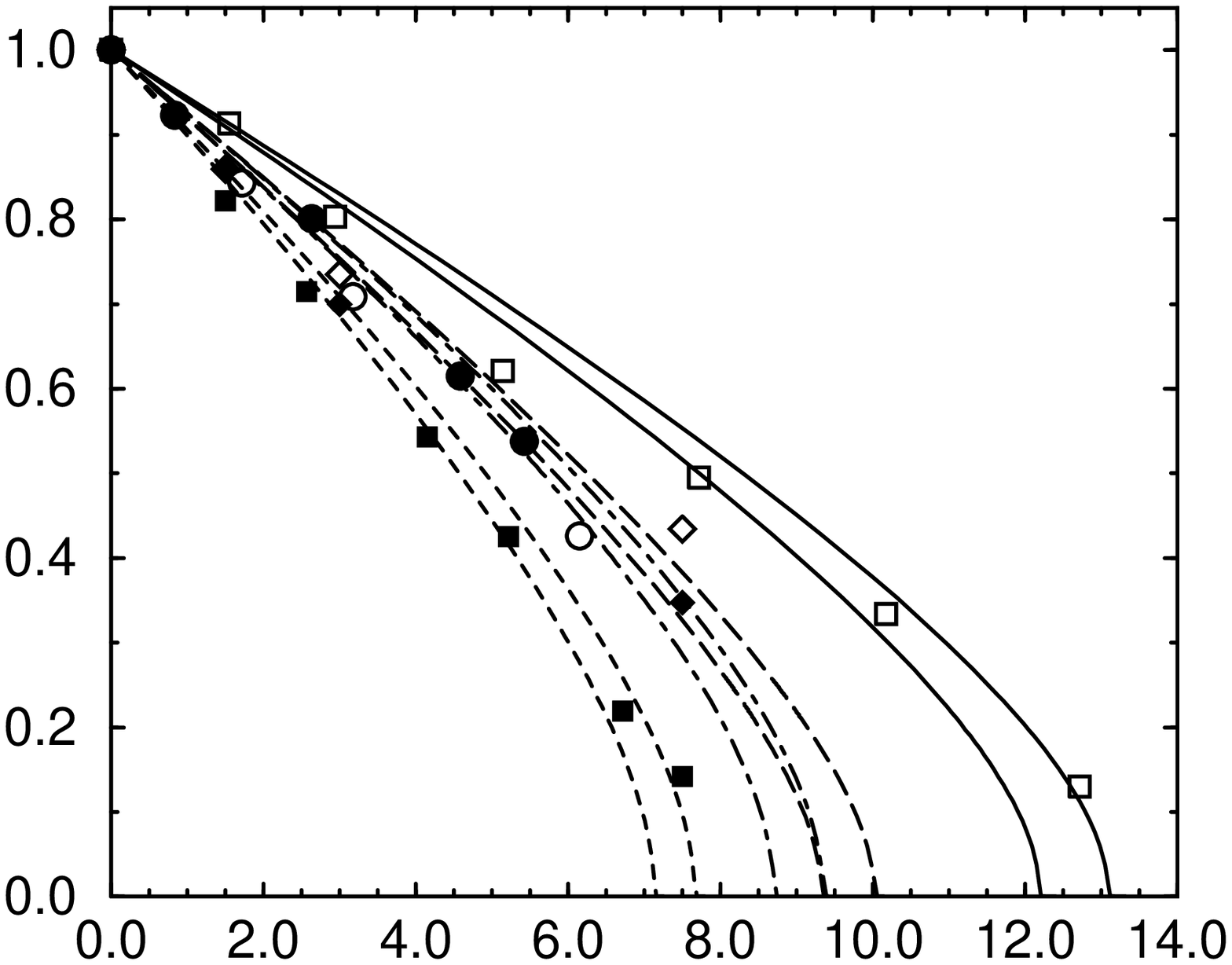,height=15cm,width=15cm} }
\parbox{0.5cm}{\hfill}
\parbox{18cm}{\LARGE\vspace{-6ex}\hfill $$n_{Zn}\;(\%\;per\;Cu\;site)
\;\;\;\;\;$$\hfill }
\caption{Fig.7:} 
\end{figure}
\end{center}

\newpage
\begin{center}
\begin{figure}[p]
\parbox{0.1cm}{\LARGE\vfill $$T_c/T_{c_{0}}$$\vspace{0.5ex}\vfill }
\parbox{15cm}{\epsfig{file=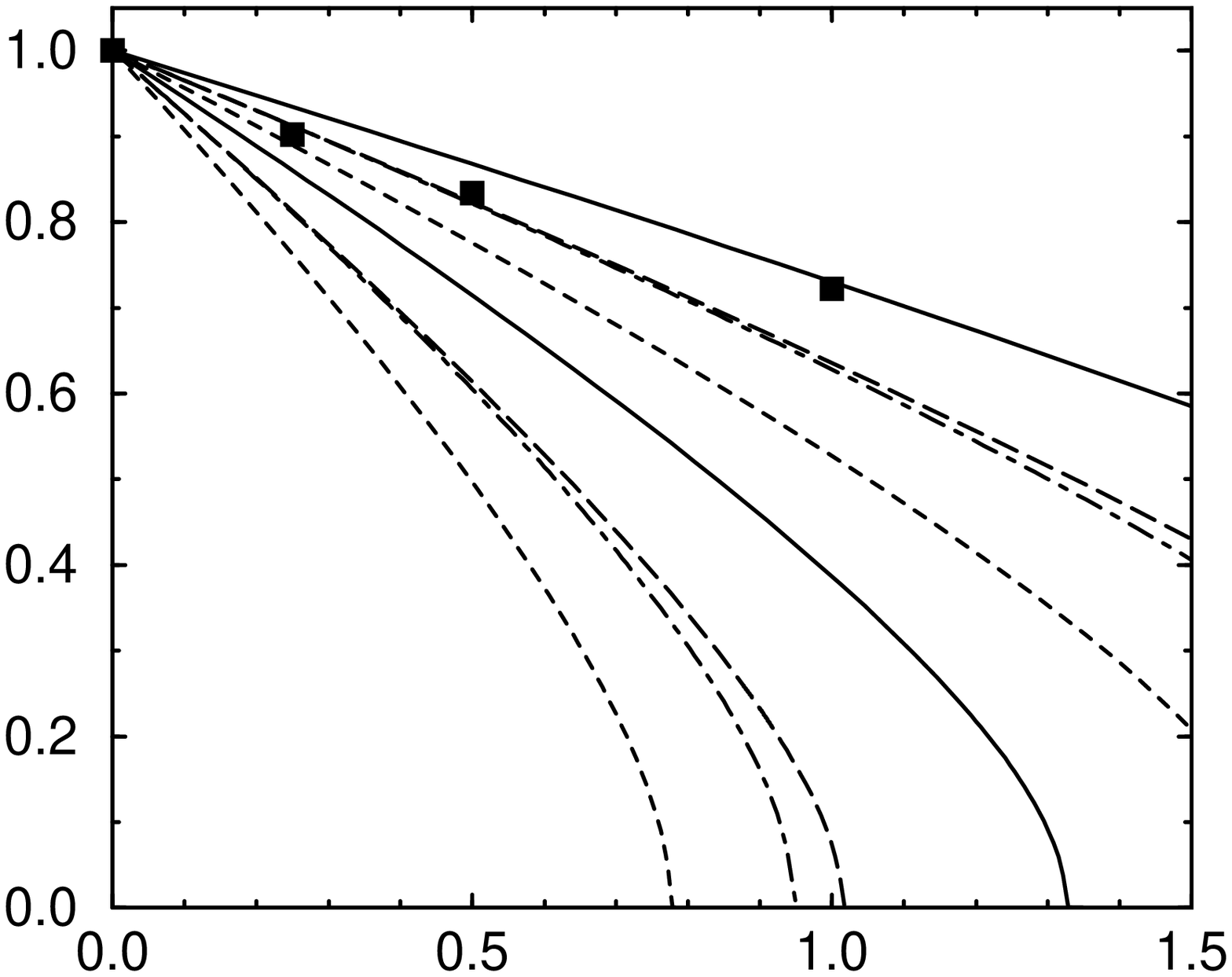,height=15cm,width=15cm} }
\parbox{0.5cm}{\hfill}
\parbox{18cm}{\LARGE\vspace{-6ex}\hfill $$n_{Zn}\;(\%\;per\;Cu\;site)
\;\;\;\;\;$$\hfill }
\caption{Fig.8a:} 
\end{figure}
\end{center}

\newpage
\begin{center}
\begin{figure}[p]
\parbox{0.1cm}{\LARGE\vfill $$T_c/T_{c_{0}}$$\vspace{0.5ex}\vfill }
\parbox{15cm}{\epsfig{file=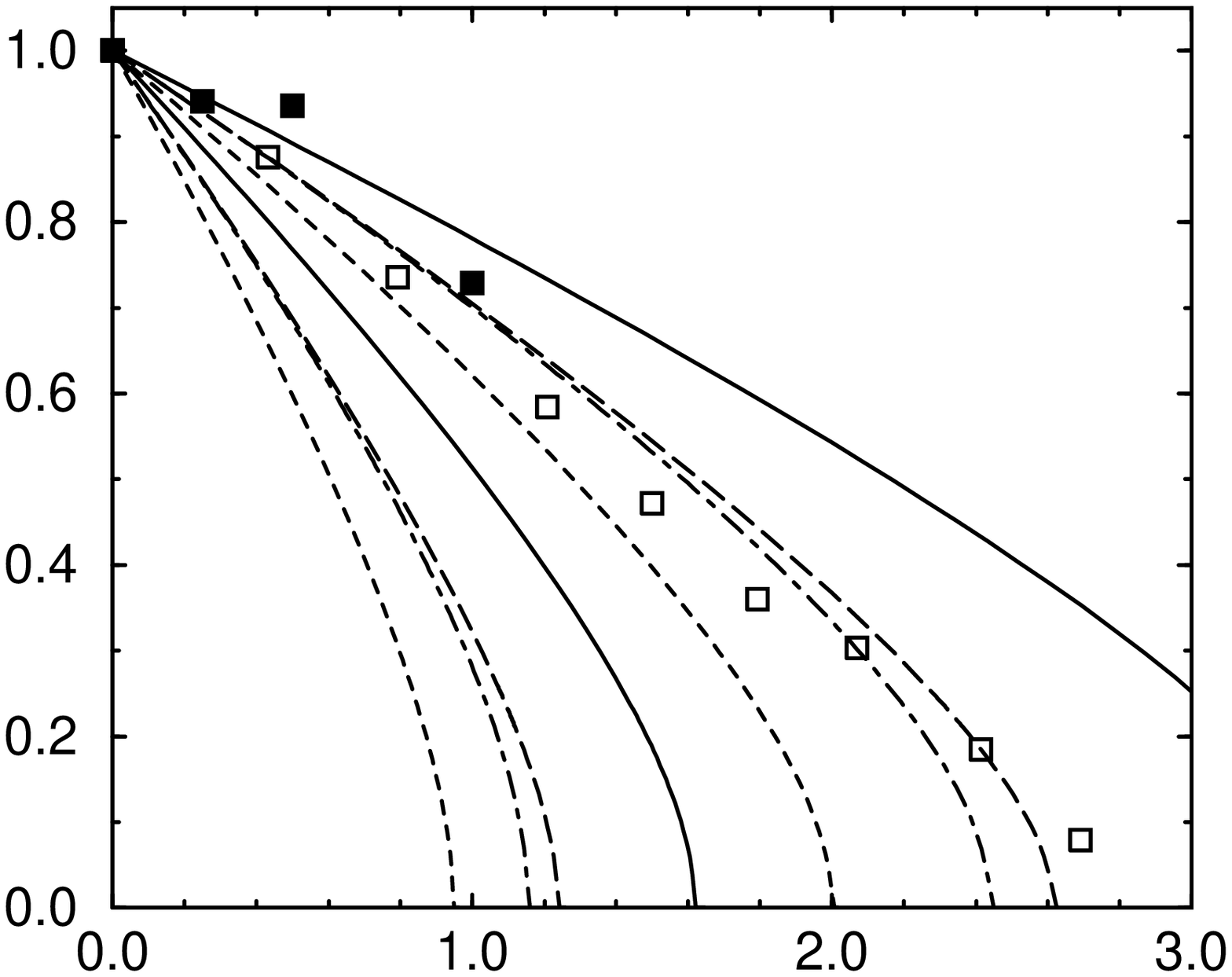,height=15cm,width=15cm} }
\parbox{0.5cm}{\hfill}
\parbox{18cm}{\LARGE\vspace{-6ex}\hfill $$n_{Zn}\;(\%\;per\;Cu\;site)
\;\;\;\;\;$$\hfill }
\caption{Fig.8b:} 
\end{figure}
\end{center}

\newpage
\begin{center}
\begin{figure}[p]
\parbox{0.1cm}{\LARGE\vfill $$T_c/T_{c_{0}}$$\vspace{0.5ex}\vfill }
\parbox{15cm}{\epsfig{file=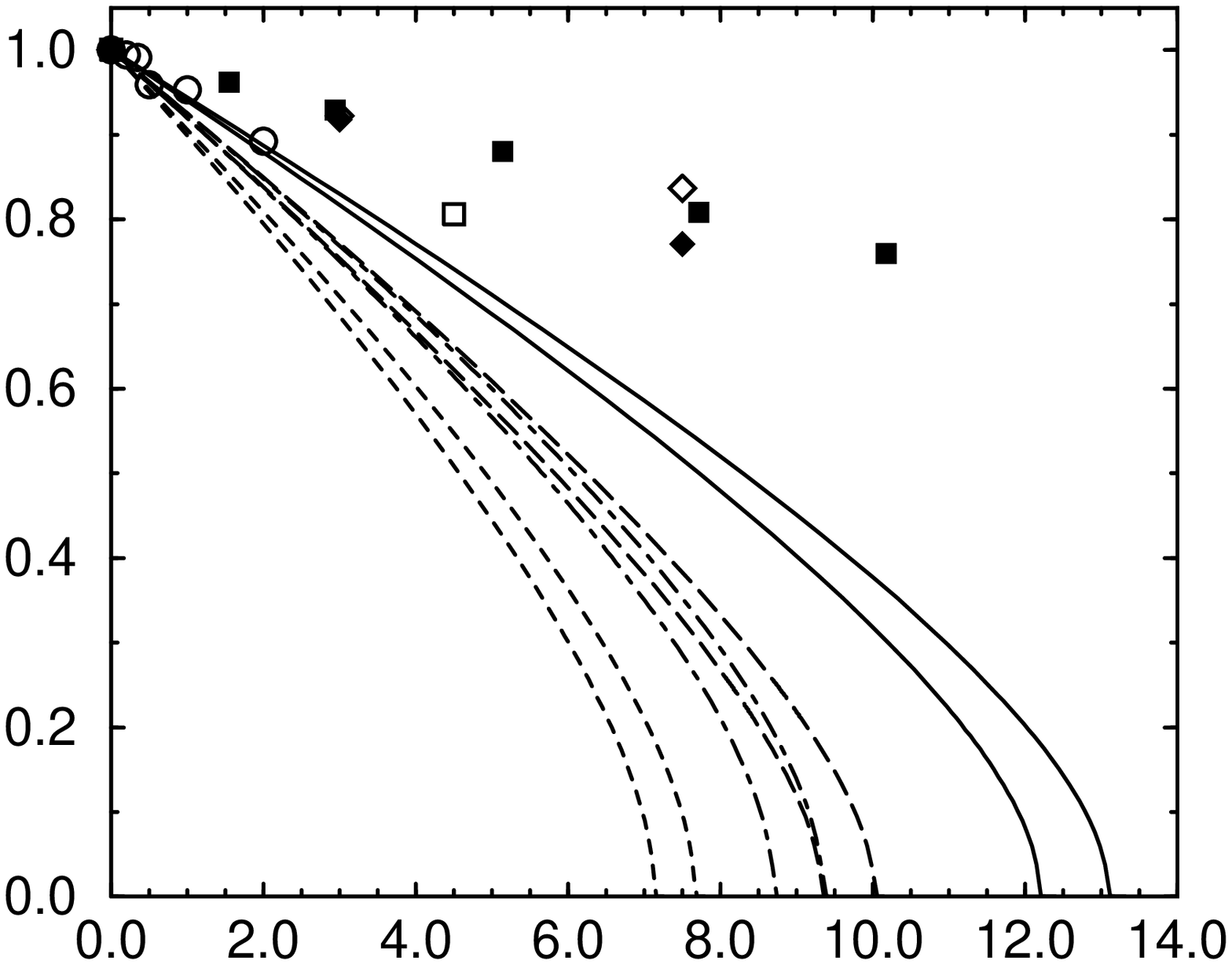,height=15cm,width=15cm} }
\parbox{0.5cm}{\hfill}
\parbox{18cm}{\LARGE\vspace{-6ex}\hfill $$n_{Ni}\;(\%\;per\;Cu\;site)
\;\;\;\;\;$$\hfill }
\caption{Fig.9:} 
\end{figure}
\end{center}

\newpage
\begin{center}
\begin{figure}[p]
\parbox{0.1cm}{\LARGE\vfill $$T_c/T_{c_{0}}$$\vspace{0.5ex}\vfill }
\parbox{15cm}{\epsfig{file=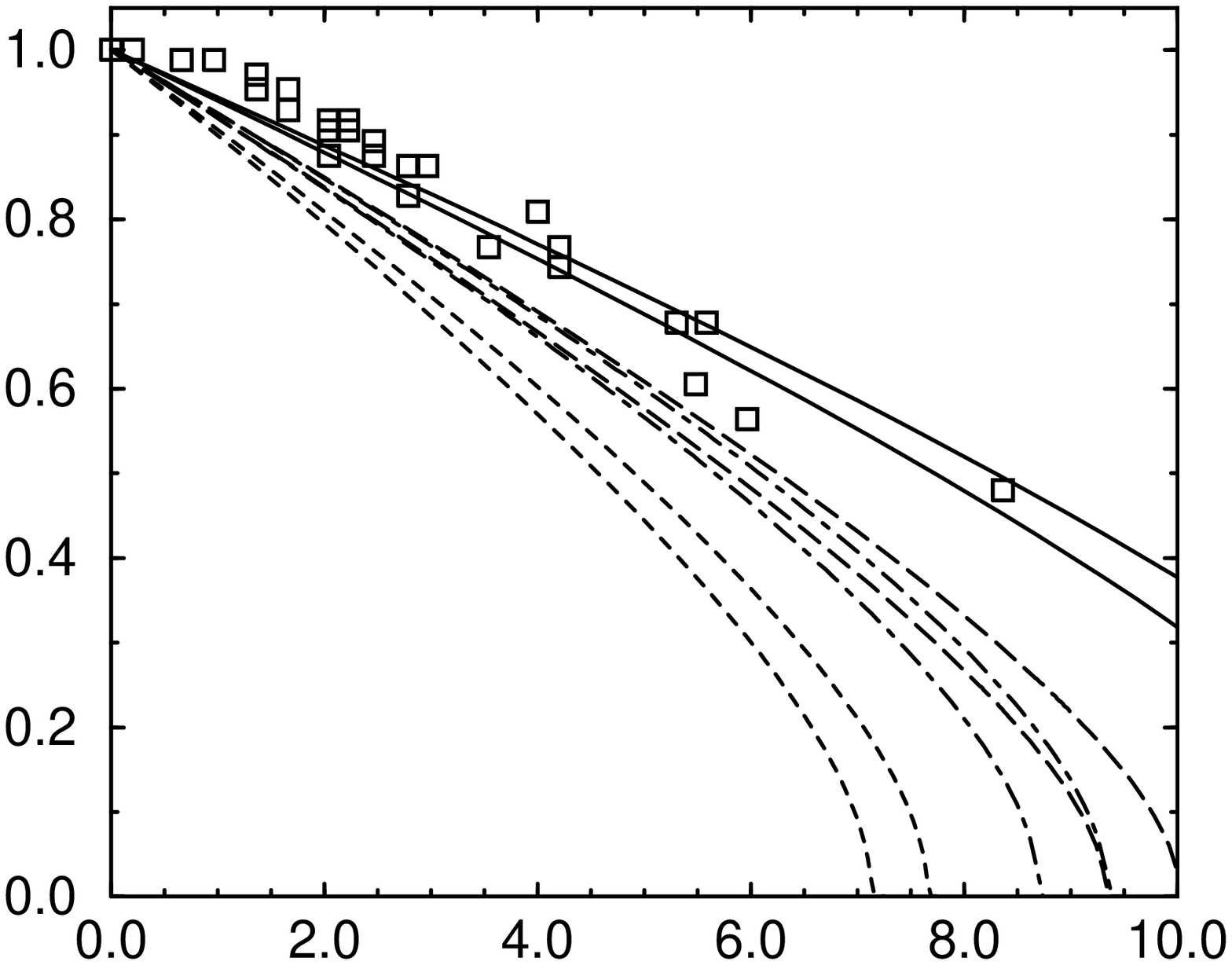,height=15cm,width=15cm} }
\parbox{0.5cm}{\hfill}
\parbox{18cm}{\LARGE\vspace{-6ex}\hfill $$n_{(O\;defects)}\;(\%\;per\;Cu\;site)
\;\;\;\;\;$$\hfill }
\caption{Fig.10:} 
\end{figure}
\end{center}


\begin{references}
\bibitem{II5a} It should be noted that the potential discussed in Refs. 33-35  
does not have this property and is virtually isotropic unless the cylindrical 
FS symmetry is broken.
\end{references}
\end{document}